\documentclass[preprint2]{aastex}
\usepackage[usenames]{color}
\usepackage{natbib}
\usepackage{hyperref}
\hypersetup{colorlinks=true, citecolor=blue}
\usepackage[yyyymmdd]{datetime}

\begin{document}
 
\title{Dust heating sources in galaxies: the case of M33 (HERM33ES)}

\author{M. Boquien\altaffilmark{1,2}, D. Calzetti\altaffilmark{1}, F. Combes\altaffilmark{3}, C. Henkel\altaffilmark{4}, F. Israel\altaffilmark{5}, C. Kramer\altaffilmark{6}, M. Rela\~{n}o\altaffilmark{7}, S. Verley\altaffilmark{7}, P. van der Werf\altaffilmark{5}, E. M. Xilouris\altaffilmark{8} \& the HERM33ES team}
\email{boquien@astro.umass.edu}

\altaffiltext{1}{University of Massachusetts, Department of Astronomy, LGRT-B 619E, Amherst, MA 01003, USA}
\altaffiltext{2}{Laboratoire d'Astrophysique de Marseille, UMR 6110 CNRS, 38 rue F. Joliot-Curie, 13388, Marseille, France}
\altaffiltext{3}{Observatoire de Paris, LERMA, 61 Av. de l'Observatoire, 75014 Paris, France}
\altaffiltext{4}{Max-Planck-Institut f\"ur Radioastronomie, Auf dem H\"ugel 69, 53121 Bonn, Germany}
\altaffiltext{5}{Leiden Observatory, Leiden University, PO Box 9513, NL 2300 RA Leiden, The Netherlands}
\altaffiltext{6}{Instituto Radioastronomia Milimetrica, Av. Divina Pastora 7, Nucleo Central, E-18012 Granada, Spain}
\altaffiltext{7}{Dept. F\'{i}sica Te\'{o}rica y del Cosmos, Universidad de Granada, Spain}
\altaffiltext{8}{Institute of Astronomy and Astrophysics, National Observatory of Athens, P. Penteli, 15236 Athens, Greece}

\begin{abstract}
Dust emission is one of the main windows to the physics of galaxies and to star formation as the radiation from young, hot stars is absorbed by the dust and reemitted at longer wavelengths. The recently launched Herschel satellite now provides a view of dust emission in the far--infrared at an unequaled resolution and quality up to 500~$\mu$m. In the context of the Herschel HERM33ES open time key project, we are studying the moderately inclined Scd local group galaxy M33 which is located only 840~kpc away. In this article, using Spitzer and Herschel data ranging from 3.6~$\mu$m to 500~$\mu$m, along with HI, H$\alpha$ maps, and GALEX ultraviolet data we have studied the emission of the dust at the high spatial resolution of 150~pc. Combining Spitzer and Herschel bands, we have provided new, inclination corrected, resolved estimators of the total infrared brightness and of the star formation rate from any combination of these bands. The study of the colors of the warm and cold dust populations shows that the temperature of the former is, at high brightness, dictated by young massive stars but, at lower brightness, heating is taken over by the evolved populations. Conversely, the temperature of the cold dust is tightly driven by the evolved stellar populations.
\end{abstract}


\section{Introduction}

The infrared (IR) is one of the main windows to the physics of galaxies. Not only does it give access to the stellar content of galaxies in the near--IR \citep{bell2001a}, it also provides us with insight into the fundamental processes of star formation. Indeed, as dust absorbs the energetic radiation from massive young stars, it reemits this energy in the IR. In other words, the IR traces star--formation.

The advent of IR telescopes such as IRAS (Infrared Astronomical Satellite) and its successors ISO (Infrared Space Observatory), Spitzer, Akari, and now WISE (Wide-field Infrared Survey Explorer) and Herschel has revolutionized our view on the universe with the discovery of a large number of dust-enshrouded galaxies whose existence had never been suspected. The mid-- and far--IR have now become a cornerstone to our understanding of galaxy formation and evolution, accounting for a third of the bolometric luminosity in quiescent galaxies such as the Milky Way, 70\% in more actively star--forming galaxies such as M82, and up to 99\% for ULIRGs \citep[Ultra Luminous Infrared Galaxies,][]{lagache2005a}.

The recently--launched Herschel Space Observatory provides a continuous coverage from 70~$\mu$m to 500~$\mu$m at an unprecedented resolution and sensitivity for the longer wavelengths. It probes dust--enshrouded star--formation across cosmic time, from the local group to high--redshift galaxies. Its exquisite resolution (from $\sim6$" at 70~$\mu$m to $\sim 38$" at 500~$\mu$m) makes it possible to resolve and study individual star--forming regions within nearby galaxies \citep[][for instance]{beirao2010a,bendo2010a,boquien2010b,cormier2010a,verley2010b}. For galaxies at cosmological distances, deep IR surveys are key tools for understanding the physics of galaxy formation and evolution \citep[][and many others]{chary2001a,dale2001a,dale2002a}. The sensitivity of Herschel allows us to probe star--formation in the IR down to the regime of quiescent star--forming galaxies at high redshift as in the case of the Herschel--GOODS deep survey. It is therefore not only timely but also of critical importance to understand the origin of the IR emission in such galaxies in the nearby universe to interpret properly these observations.

Dust emission in galaxies can have several unrelated origins as for instance suggested by \cite{sauvage1994a,kewley2002a}. (1) Massive stars in star--forming regions heat and excite the dust, as described above. This emission is generally compact and corresponds to clumps of star--formation within galaxies. (2) A large scale diffuse emission may arise due to cirrus \citep{helou1986a} being heated by (a) energetic radiation escaping from individual star--forming regions \citep{sauvage1990a,xu1990a,popescu2000a,misiriotis2001a,popescu2002a,popescu2005a,tabatabaei2007a} and/or (b) the general radiation field of evolved stars \citep{helou1986a,xu1996a,li2002a,boselli2004a}. (3) Hot grains in the photosphere or circumstellar atmosphere of mass--losing stars also contribute, especially in the mid--IR \citep[][for instance]{jura1987a,knapp1992a,mazzei1994a}. (4) Embedded AGN affect the surrounding dust \citep{degrijp1985a,wu2007a}.

Observations have shown that the dust forms as soon as metals are available leading to some absorption by the dust at relatively high redshift \citep{meurer1997a,giavalisco2004a,bouwens2009a}. This means that dust also reprocesses the radiation from young stars at cosmological distances and, as a consequence, IR emission acts as a star--formation tracer even at high redshift. This hints at the fundamental importance of the IR not only to study star--formation in nearby galaxies but also in the distant universe. For a decade, studies have used the IR luminosity function to explore the formation and evolution of galaxies, providing an extinction--free estimate by essence of the cosmic star formation rate density. It serves as a test bench for models of structure formation and evolution \citep[e.g.,][]{kay2002a}.

The use of the IR as a SFR (star formation rate) estimator is based on three assumptions. First, it is assumed that dust opacity is infinite throughout the galaxy. That is, all the energetic radiation from young stellar populations is absorbed and reemitted in the IR. While this assumption is fulfilled in LIRGs and ULIRGs, in more quiescent galaxies it is not the case. The combination of the IR continuum luminosity with an optical recombination line or the UV (ultraviolet) luminosity permits to circumvent this problem \citep{calzetti2007a,kennicutt2007a,leroy2008a,bigiel2008a,kennicutt2009a}. The second assumption is that the young stellar populations dominate the radiation field in extinction--sensitive bands (UV and optical). In other words it means that AGN heating as well as heating from evolved stellar populations and dust emission in circumstellar envelopes are small compared to it. This question is of crucial importance as it determines whether a given IR band is a reliable star--formation tracer or not. As the diffuse emission can represent up to nearly 90\% of the IR flux in Sa galaxies \citep{sauvage1992a}, it can lead to an error on the SFR up to an order of magnitude depending on its origin. Finally the last assumption is that the initial mass function is universal.

There is an active and on--going debate in the literature regarding the actual process giving rise to the diffuse emission. Its presence has been unambiguously detected in numerous studies since \cite{lonsdale1987a}. The non star--forming related contamination is non--negligible for whole galaxies according to some authors, however its amplitude is not assessed precisely, with some studies on individual galaxies and samples of galaxies finding conversely that star--formation drives the majority of the IR luminosity \citep{sauvage1992a,devereux1994a,devereux1994b,devereux1995a,buat1996a,israel1996a,walterbos1996a,devereux1997a,jones2002a,kewley2002a,bell2003a,boselli2004a,forster2004a,hinz2004a,cannon2006b,perez2006a,calzetti2007a,montalto2009a,dacunha2010a,calzetti2010a}. These results hint at a variation from galaxy to galaxy in a more fundamental way than the relation with galaxy type found by \cite{sauvage1992a}. When studying individual star--forming regions, it appears that the warm dust peaking around 70~$\mu$m is a good tracer of star formation as was found both in the Magellanic Clouds \citep{lawton2010a} and in star forming regions in nearby galaxies \citep{li2010a}.

Here we use Herschel observations taken in the context of the HERM33ES open time key project \citep{kramer2010a} in combination with Spitzer IRAC (Infrared Array Camera) and MIPS (Multiband Imaging Photometer for Spitzer) data as well as GALEX FUV (far--ultraviolet), and ground--based H$\alpha$ maps of a local group galaxy, in order to characterize the emission from the warm and cold dust populations\footnote{The case of the PAH properties will be treated extensively in Rosolowsky et al. (in preparation). We concentrate here primarily on the emission of the warm and cold dust components.}. By convention, we designate as ``warm dust'' the population that peaks typically at $\mathrm{\sim50~K}$ and by ``cold dust'' the one that peaks at $\mathrm{\sim20~K}$. The selected target is M33, a nearby Scd galaxy located only 840~kpc away \citep{freedman1991a} with an inclination of 56$^{\circ}$ \citep{regan1994a}. It provides us with an exceptional physical resolution better than $\sim150$~pc at 500~$\mu$m allowing us to resolve the various morphological components of the galaxy. Finally, the absence of an active nucleus excludes one possible dust heating mechanism and a shallow metallicity gradient limits the influence of the radial evolution of the metallicity on the emission of the dust. 

In Sec.~\ref{sec:obs} we present the observations and the  data reduction. Sec.~\ref{sec:results} provides the results, which are discussed in Sec.~\ref{sec:discussion}. The main conclusions are given in Sec.~\ref{sec:conclusion}.

\section{Observations and data processing}
\label{sec:obs}
\subsection{Observations}

The key observations we use here cover the entire body of M33 from the near--IR to the far--IR from 3.6~$\mu$m to 500~$\mu$m. They trace the stellar mass, the PAHs (Polycyclic Aromatic Hydrocarbons), the VSGs (Very Small Grains), and the BGs (Big Grains). We combine observations made with Spitzer IRAC at 3.6~$\mu$m and 8.0~$\mu$m and MIPS at 24~$\mu$m and 70~$\mu$m \citep{verley2007a} with Herschel PACS (Photodetector Array Camera and Spectrometer) and SPIRE (Spectral and Photometric Imaging Receiver) observations obtained in the context of the HERM33ES open time key project \citep{kramer2010a}. PACS covers the 100~$\mu$m and 160~$\mu$m bands and SPIRE covers the 250~$\mu$m, 350~$\mu$m, and 500~$\mu$m bands. Herschel observations were carried out on 2010--01--07 for a total duration of 6.3 hours at a 20\arcsec/s scanning speed in parallel mode with one scan in each perpendicular direction. They cover the entire galaxy including the outermost regions.

We also use ground--based H$\alpha$ observations presented in \cite{hoopes2000a} in order to trace the current ($\sim10$~Myr) star formation and  GALEX data \citep{gildepaz2007a} to trace the recent ($\sim100$~Myr) star formation. Finally, we use the HI maps published in \cite{gratier2010a} to trace the neutral gas. A complete mapping of M33 in CO is being completed to probe the molecular component and will be presented in Braine et al. (in preparation).

\subsection{Data processing}

\subsubsection{Herschel}

We have processed Herschel data using two different methods to improve products over the standard automatic pipeline which does not give satisfactory results for extended objects.

First, PACS data processing was done using HIPE version 3.0 \citep{ott2010a}. In a first step we processed the frames to level 1, flagging known bad pixels, checking and masking saturated pixels, adding pointing information to each pixel, and calibrating the frames. We have then corrected the frames for the natural drift of the signal of each bolometer during the scan to retrieve the emission. To do so, for each scanleg, we have fitted a linear function for each bolometer on the first and last 10\% of the readouts, which cover only the sky, and subtracted this function from the signal. To eliminate the glitches, we have joined the scan and the cross--scan frames and used the 2nd order deglitcher with a 5--$\sigma$ threshold. To eliminate artifacts caused by the deglitching, we have masked all readouts over a correlation length for each pixel containing a glitch. Finally, to remove the intrinsic 1/f noise from the bolometers we have made use of madmap. The absolute calibration uncertainty is better than 10\% at 100~$\mu$m and better than 20\% at 160~$\mu$m\footnote{PACS Scan Map release note version 1.1.}.

We have also processed PACS data with Scanamorphos version 1 (Roussel 2010, submitted) to ensure that the data processing method does not affect the results. Scanamorphos produces images with a background much flatter on the large scale compared to the ones produced with madmap, which is important for very large apertures. In our case the comparison of the background--subtracted fluxes obtained in each pixel for lower resolution images (Sect.~\ref{ssec:convol}) shows that madmap maps present a 10 to 20\% stronger flux, most likely due to the varying background which is much stronger at the location of the galaxy, and that the relative scatter of the ratio of madmap and scanamorphos fluxes is non negligible (31\% at 100~$\mu$m and 23\% at 160~$\mu$m, mostly concentrated on lower fluxes). The easier determination of the actual background in Scanamorphos images has led our choice to rely on these images throughout the paper.

The SPIRE data processing was also done applying HIPE version 3.0 using the standard pipeline. The data processing is described in greater detail in \cite{kramer2010a}. The typical calibration uncertainties of the SPIRE bands are about 15\%.

\subsubsection{HI, H\texorpdfstring{$\alpha$}{alpha}, GALEX and Spitzer}

The HI, H$\alpha$, GALEX and Spitzer data we use in this article were already processed and obtained from the literature \citep{hoopes2000a,gildepaz2007a,verley2007a,tabatabaei2007a,verley2009a,gratier2010a} and NED (NASA Extragalactic Database). No further processing was done besides removing some bright galactic foreground stars and convolving the images to lower resolution and registering them to a common frame (Sec.~\ref{ssec:convol}).

\subsection{Convolution to lower resolution}
\label{ssec:convol}
The aim of data processing is to produce maps at each wavelength that can be directly compared to each other. To do so, the maps need to have both the same resolution (i.e, the same PSF [Point Spread Function]), the same pixel scale and the same coordinates. We have proceeded in the following way: 1. We have converted all Herschel, Spitzer and GALEX images to flux units (Jy/pixel). For SPIRE maps that are originally in Jy/beam, we have assumed the beam area is 426\arcsec$^2$, 771\arcsec$^2$ and 1626\arcsec$^2$ at 250, 350, and 500~$\mu$m respectively\footnote{Values obtained from the SPIRE beam release notes 1.0: \url{ftp://ftp.sciops.esa.int/pub/hsc-calibration/SPIRE/PHOT/Beams_v1.0/beam_release_note_v1-0.pdf}}. To convert to flux densities, we have used the following relation: $F=f\times p^2/S$, where $F$ is the flux in Jansky, $f$ the flux in Jy/beam, $p$ the pixel size in arcsec, and $S$ the area of the beam in arcsec$^2$. 2. We have convolved all the images to the resolution of the SPIRE 500~$\mu$m image, using either the instrument PSF if the resolution difference was large or the dedicated convolution kernels provided by \cite{gordon2008a}\footnote{\url{http://dirty.as.arizona.edu/\textasciitilde kgordon/mips/conv\_psfs/conv\_psfs.html}} otherwise. 3. we have created reference frames of the desired field--of--view with a pixel size designed to be slightly larger than the PSF FWHM (Full Width Half Maximum). Several of these frames have been produced with different pixel scales, we have finally adopted frames that have a pixel size of 42\arcsec\ which is slightly larger than the resolution of SPIRE~500~$\mu$m to enclose the PSF. 4. We have registered the convolved images on the reference frame with 42\arcsec\ pixels, using {\sc iraf}'s {\sc wregister} procedure. 5. Finally we have converted all Herschel, Spitzer, and GALEX images to W~kpc$^{-2}$. The HI map has been converted to M$_\sun$~kpc$^{-2}$, and the H$\alpha$ map has been converted to W~kpc$^{-2}$. The use of surface brightness units allows for an easier comparison with other galaxies as it is distance independent. The final maps are presented in Sec.~\ref{ssec:maps}.

\subsection{Flux measurement}
\label{ssec:flux-measurement}
For the pixel--to--pixel analysis we consider each pixel in the convolved, registered image generated in Sect.~\ref{ssec:convol}, from which we have subtracted the background of the image. The background is measured averaging the mean value in $5\times5$ pixels apertures around the galaxy at an inclination--corrected distance of at least 5~kpc from the center, in the low resolution registered image. The measurement is made interactively using the {\sc imexamine} procedure in {\sc iraf}. The uncertainty is taken as the quadratic mean of the standard deviation of the background level and the mean of the pixel--to--pixel standard deviation measured in $5\times5$ pixels apertures using the same method as described above.

We have to note that due to the registration and the convolution of the images, it is difficult to estimate the uncertainties in a precise fashion relying on the noise statistics. In the final images the uncertainty is made of two identifiable components. First, the background is not completely flat on the large scale which induces a systematic uncertainty on the background level itself. While the effect is very small for bright pixels, it is much more important for fainter ones. This may be due to instrumental/data reduction artifacts as in the case of PACS data processed with madmap, which has led to the adoption of maps processed with Scanamorphos. It may also be due to the presence of foreground cirrus in the IR which are clearly seen in the SPIRE bands. Another source of uncertainty is the presence of compact sources, such as foreground stars in the Milky Way, which is particularly problematic in the 3.6~$\mu$m band. These two noise sources contribute generally to the same order of magnitude to the uncertainty. Knowing these limitations, we assume that the standard deviation measured as explained above provides a good estimate of 1--$\sigma$ error bars. Pointing uncertainties should have a limited effect as they are small compared to the final pixel size.

For the GALEX FUV and H$\alpha$ images, the fluxes have additionally been corrected for the foreground galactic extinction using the \cite{cardelli1989a} law, assuming $\mathrm{E(B-V)=0.042}$, obtained from NED. The H$\alpha$ fluxes have also been corrected for the NII line contamination assuming [NII]/H$\alpha$=0.05 within the H$\alpha$ narrow--band filter \citep{hoopes2000a}.

\section{Results}
\label{sec:results}

In this section we present and  describe individual band maps and color maps. We have selected pixels that have a SNR (Signal--to--Noise Ratio) of at least 3 in both bands for the color maps. These pixels are located within 5~kpc from the center of the galaxy.

\subsection{Individual bands}
\label{ssec:maps}
\begin{center}
\begin{figure*}[!htbp]
\includegraphics[width=\textwidth]{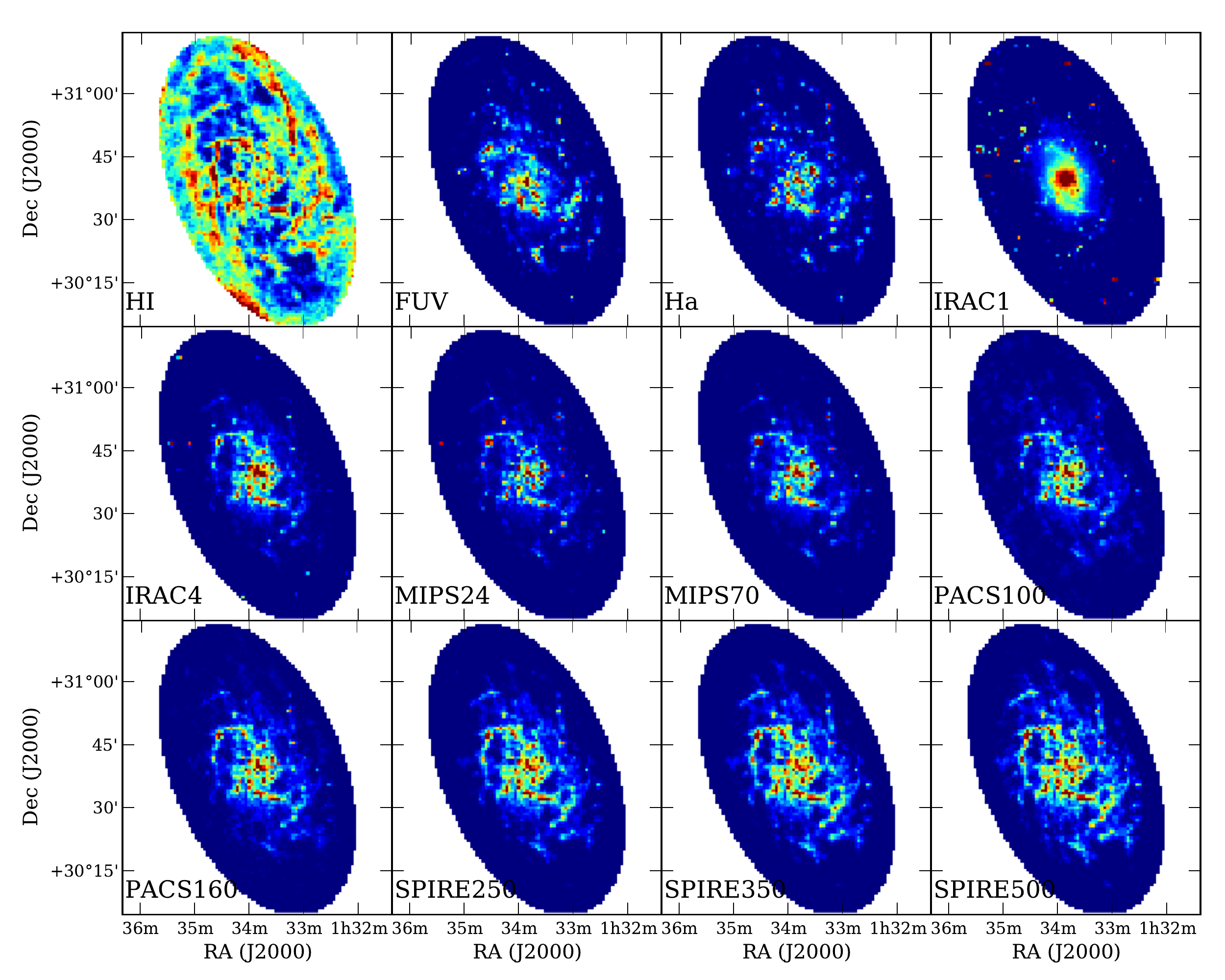}
\caption{From top--left to bottom--right, maps of M33 in HI, FUV, H$\alpha$, 3.6, 8.0, 24, 70, 100, 160, 250, 350, and 500~$\mu$m. Blue indicates the lowest value and red the highest one, in linear scale, eliminating the 2.5\% faintest and brightest pixels. The distance from the center is limited to 5~kpc, represented by the ellipse. All images have been convolved to the SPIRE 500~$\mu$m resolution and have been registered to a common frame with 42\arcsec\ pixels.\label{fig:maps}}
\end{figure*}
\end{center}

In Fig.~\ref{fig:maps} we present maps of M33 in individual bands convolved to the SPIRE 500~$\mu$m PSF and registered to a common reference frame with 42\arcsec\ pixels as described in Sec.~\ref{ssec:convol}.

These maps trace the neutral atomic gas reservoir (HI), recent and current star formation (FUV and H$\alpha$), the evolved stellar populations (3.6~$\mu$m), the PAHs (8.0~$\mu$m), the warm dust (24~$\mu$m to 160~$\mu$m), and the cold dust (250~$\mu$m to 500~$\mu$m). Note that depending on the radiation field intensity, the 160~$\mu$m band may be dominated either by the warm or the cold dust \citep{bendo2010a}. For convenience in this paper we associate the 160~$\mu$m band with the 70~$\mu$m and 100~$\mu$m bands that are dominated by the warm dust.

The spiral structure is unsurprisingly marked in all bands except for the 3.6~$\mu$m band in which the emission is dominated by evolved stellar populations throughout the disk. This band shows a markedly different structure, with the spiral arms being relatively faint and the surface brightness declining smoothly from the inner to the outer regions.

The neutral atomic gas as traced by HI exhibits a spiral structure extending to a larger distance than what is seen in bands tracing star formation. It presents several regions that seem to be almost entirely depleted of atomic gas. These regions show very little emission in dust tracing bands and also in H$\alpha$ and FUV. On average however, the azimuthally averaged neutral atomic gas column density varies little with radius up to 8~kpc from the center of the galaxy \citep{verley2009a,gratier2010a}.

Faint emission is detected farther from the center in SPIRE bands from 250~$\mu$m to 500~$\mu$m compared to what is seen in PACS bands at 100~$\mu$m and 160~$\mu$m. This may be due to the relative shallowness of the PACS maps as this diffuse emission is more marked in the MIPS~160 band. However when comparing the convolved, registered PACS~160 and MIPS~160 images, they yield similar fluxes for faint regions, albeit with a relatively large scatter.

The massive star forming region to the North-East, NGC~604, appears clearly and is particularly prominent in bands tracing the warm dust. There is an excess of emission in the 3.6~$\mu$m band (and also in the 4.5~$\mu$m band which is not presented here), maybe due to emission of nebular lines or continuum, and the 3.3~$\mu$m PAH line as well as the continuum from very hot dust and transiently heated grains. This excess has already been observed in those bands in strongly star--forming regions in other galaxies and interacting systems \citep{smith2009a,mentuch2009a,boquien2010c}.

\subsection{Old and young stellar populations}
\label{ssec:IRAC1}

Being sensitive to the Rayleigh-Jeans tail of the photospheric emission of stars, the IRAC 3.6~$\mu$m band is a tracer of the stellar mass in the galaxy. Though it can be affected by the PAH 3.3~$\mu$m line, nebular emission and hot dust in HII regions as well as VSGs fluctuating to temperatures of several hundreds K, such as in NGC~604 where it is not as reliable as a tracer of the stellar mass.

Normalized to the 3.6~$\mu$m band, the FUV and H$\alpha$ bands show a tenuous spiral structure and do not present any striking radial structure, showing that on the large scale, unextinguished star formation varies similarly to the stellar mass. Local clumps of star formation that depart from this trend are clearly seen, corresponding to star forming regions in spiral arms.



\subsection{Warm dust}
\label{ssec:warm}

\begin{center}
\begin{figure*}[!htbp]
\includegraphics[width=\textwidth]{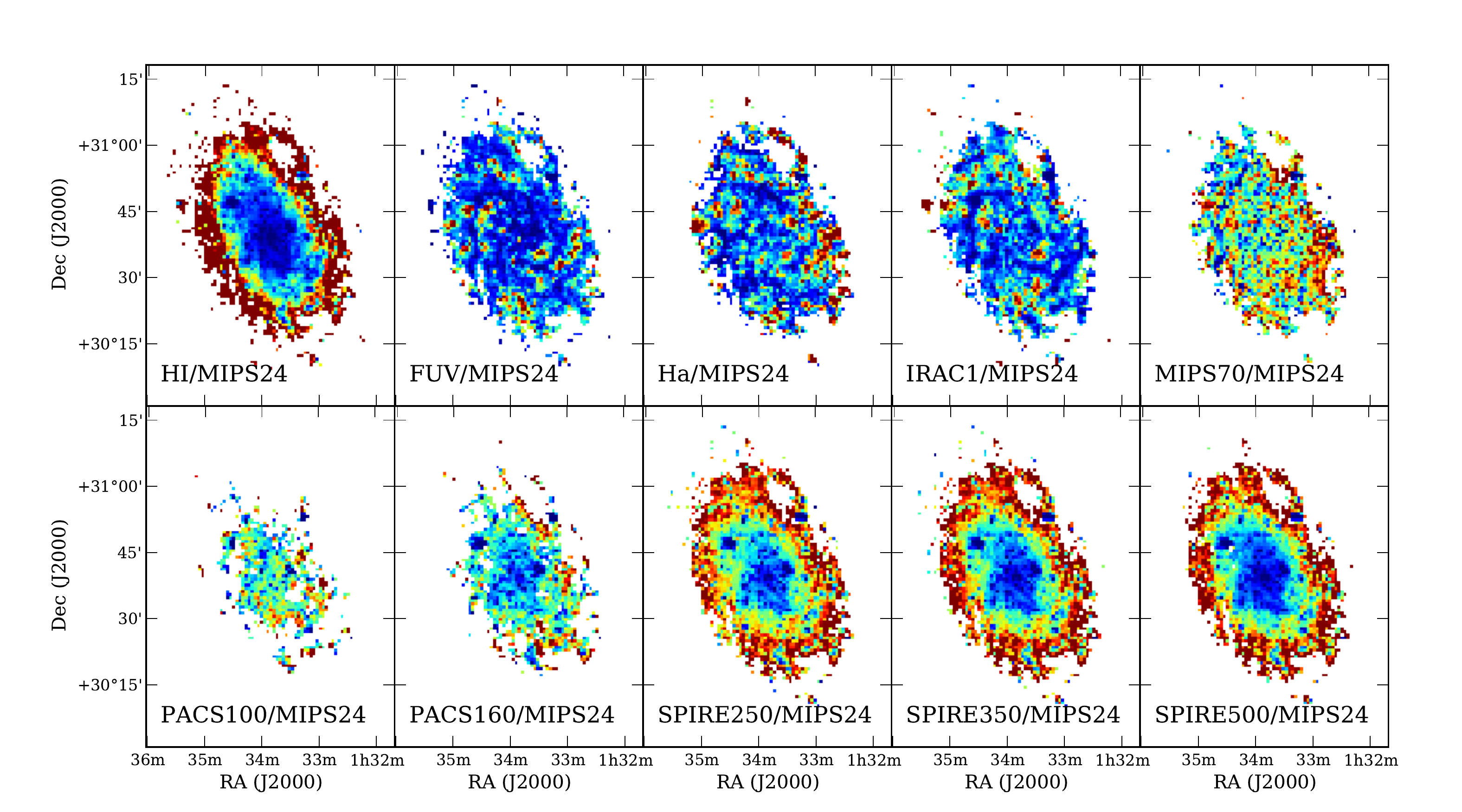}
\includegraphics[width=\textwidth]{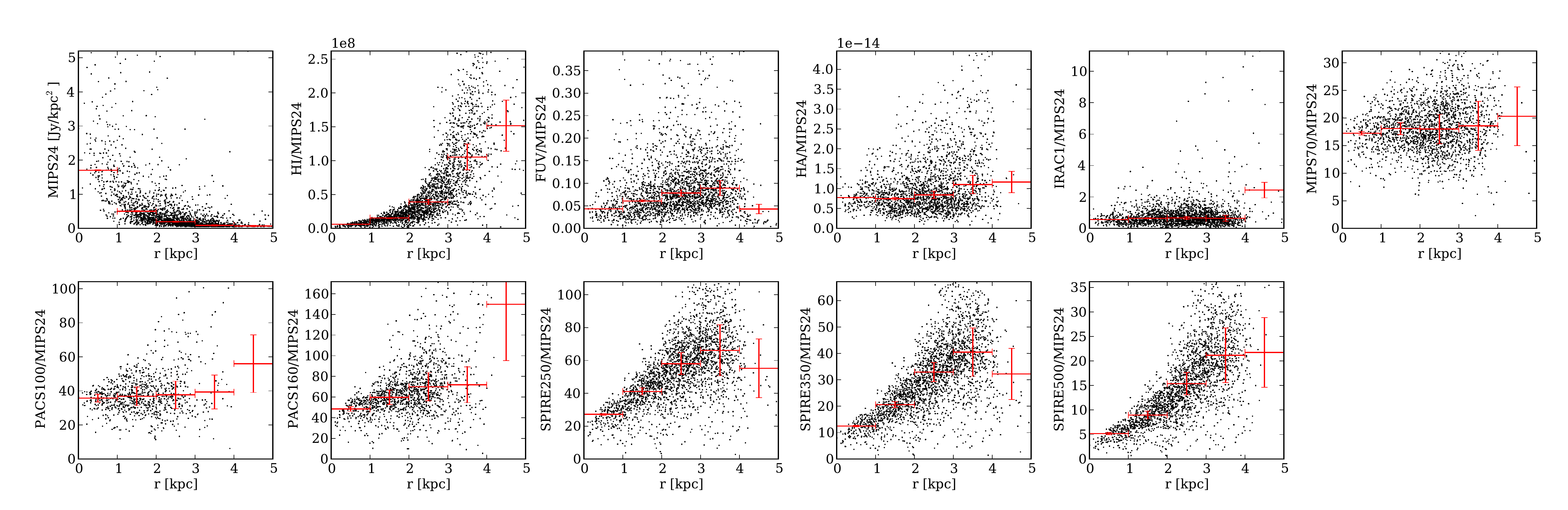}
\includegraphics[width=\textwidth]{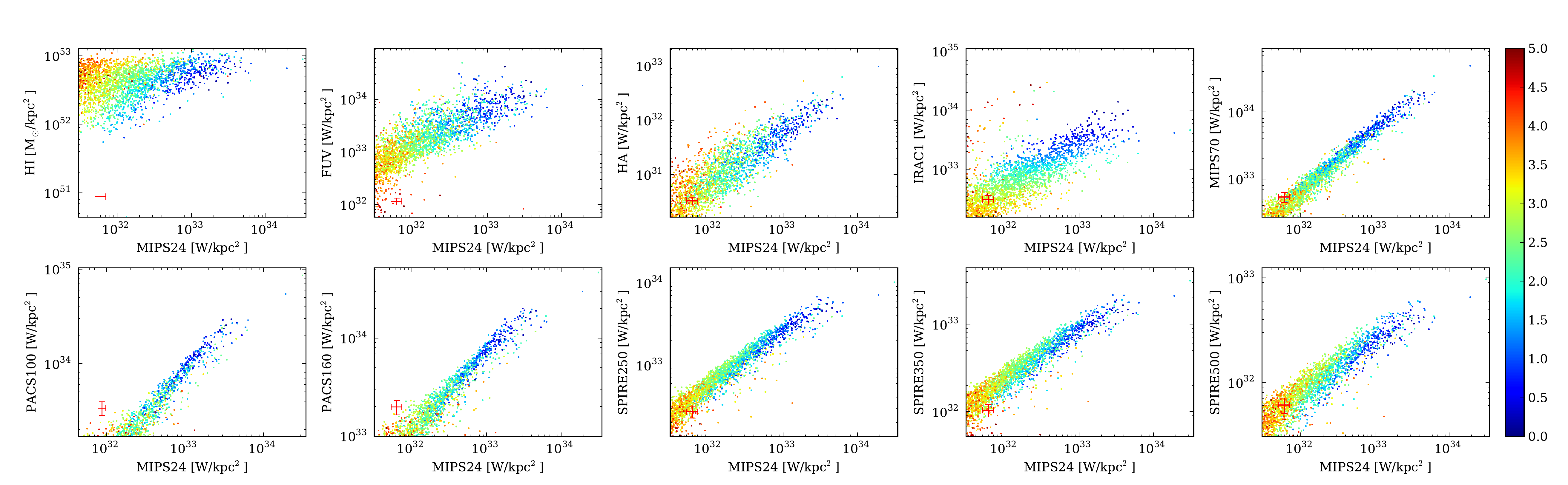}
\caption{Top: color maps between all the bands and the 24~$\mu$m band surface brightness. Blue indicates a low ratio and red indicates a high ratio. Middle: color between all the bands and the 24~$\mu$m band surface brightness as a function of the radial distance. The error bars in red represent the median 1--$\sigma$ uncertainty in bins that have a width of 1~kpc. Bottom: brightness of all the bands as a function of the 24~$\mu$m band surface brightness. The color indicates the galactocentric distance, following the colorbar at the bottom--right corner in units of kpc, blue for regions close to the center and red for regions located in the outskirts at 5~kpc. The error bars in red represent the 1--$\sigma$ uncertainty.\label{fig:MIPS24}}
\end{figure*}
\end{center}

\begin{center}
\begin{figure*}[!htbp]
\includegraphics[width=\textwidth]{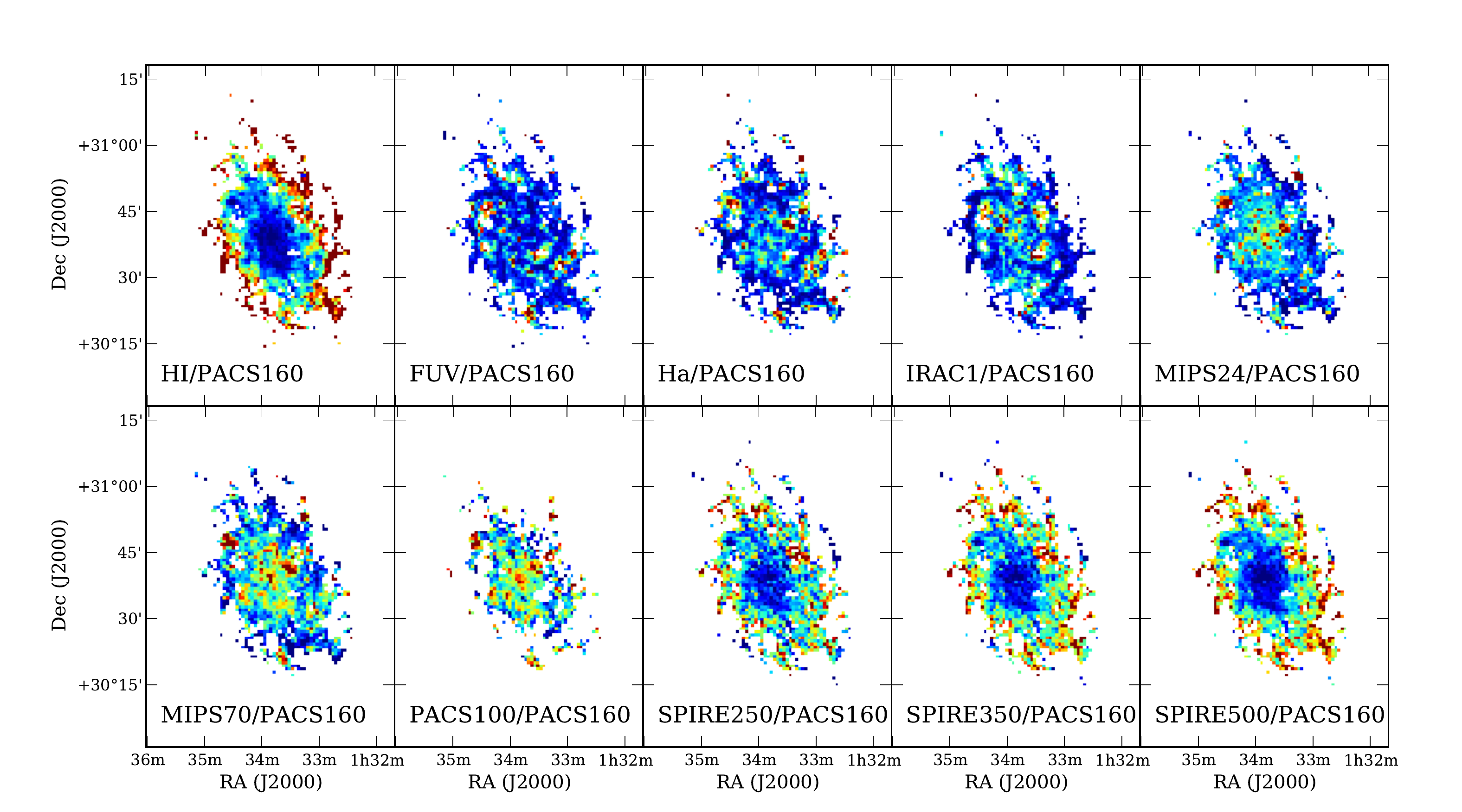}
\includegraphics[width=\textwidth]{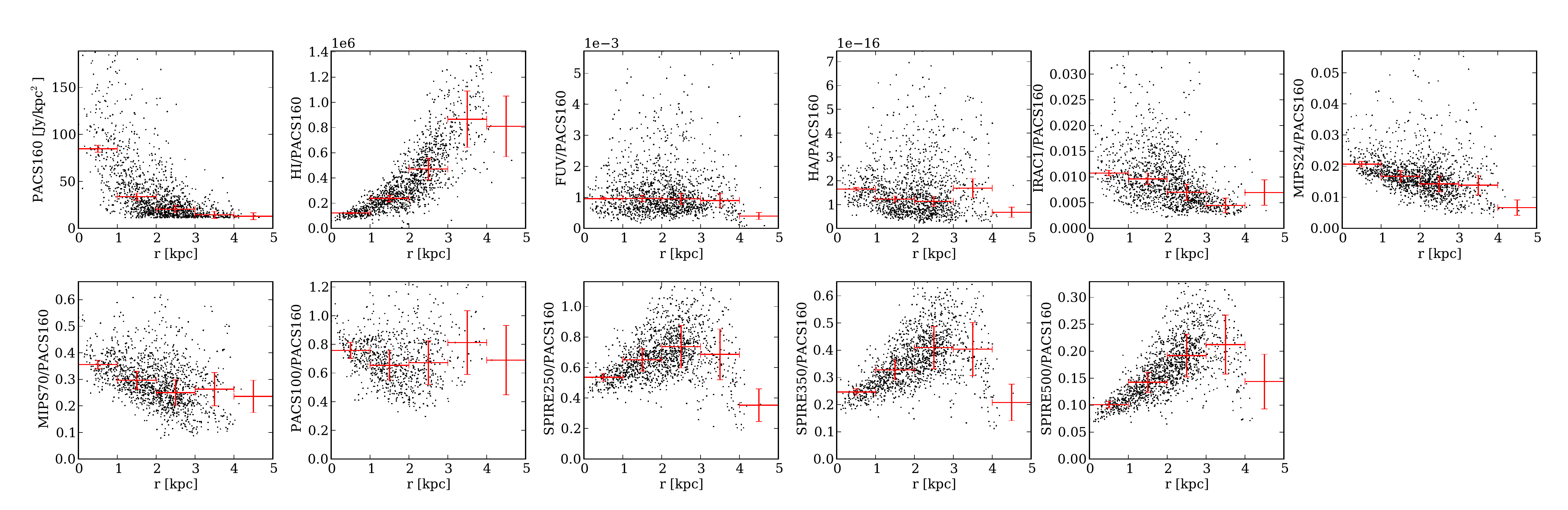}
\includegraphics[width=\textwidth]{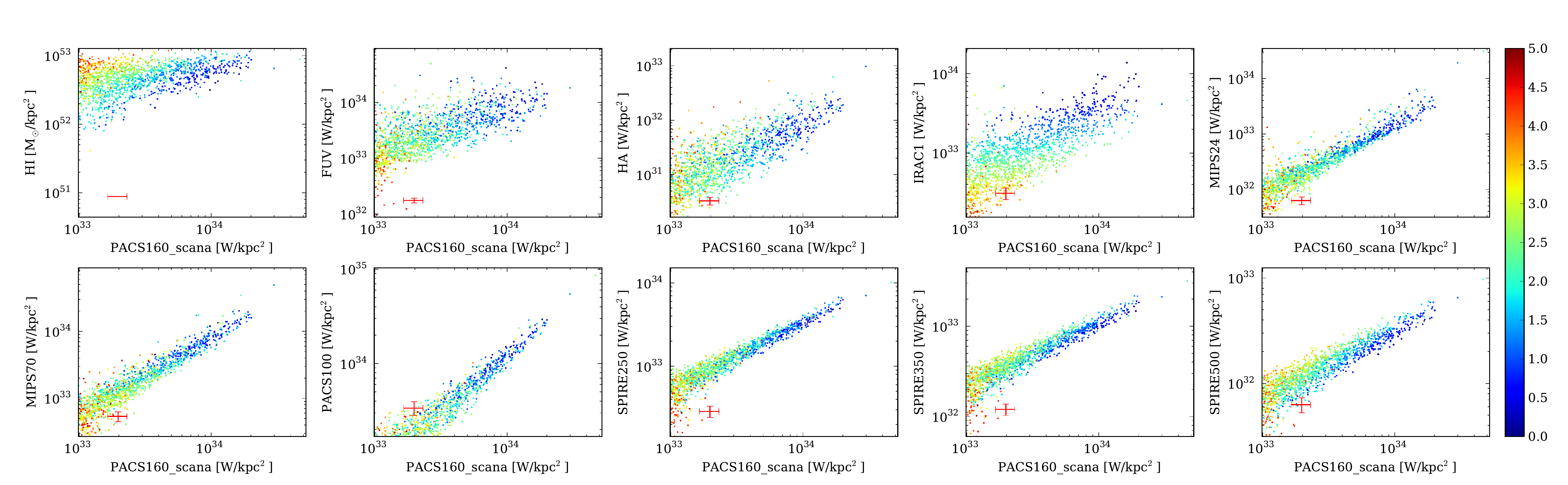}
\caption{Same as in Fig.~\ref{fig:MIPS24} but for the 160~$\mu$m band surface brightness.\label{fig:PACS160}}
\end{figure*}
\end{center}

The bands tracing the warm dust span the peak of the FIR emission and as such are major tracers of the TIR (total infrared) emission in the galaxy. The warm dust emission is as expected clearly enhanced within spiral arms compared to the stellar mass as we can see at 24~$\mu$m and 160~$\mu$m in Fig.~\ref{fig:MIPS24} and \ref{fig:PACS160}.

However, there are several regions that appear to contain a high stellar mass compared to the intensity of the warm dust emission. Actually these correspond to holes in the galaxy containing very little neutral gas, and hence very little dust. If we compare bands tracing the warm dust, we get an indication of the temperature. Generally we find a slight radial trend from higher temperatures in the inner regions towards colder temperatures in the outer regions. However if we look at the 70/100 color, we do not see any significant trend with the radial distance and the ratio indicates that the peak is longward of 70~$\mu$m in terms of $\mathrm{F_\nu}$. However, there is a weak, decreasing, radial trend when comparing the 70~$\mu$m and 100~$\mu$m to the 160~$\mu$m emission, albeit with a large scatter. Comparing to bands tracing the cold dust (250~$\mu$m to 500~$\mu$m), there is a radial gradient, the warm dust emission decreasing relatively to the cold dust when going outward. This gradient is more pronounced when comparing shorter bands, especially those at 24~$\mu$m and 70~$\mu$m, to the cold dust. The scatter increases as we compare bands that are far apart. This is probably due to the fact that these bands are increasingly dominated by different dust populations and heating mechanisms. We also observe a trend with the radial distance: at a given warm dust brightness, the cold dust brightness tends to be higher in the outer regions.

\subsection{Cold dust}

\begin{center}
\begin{figure*}[!htbp]
\includegraphics[width=\textwidth]{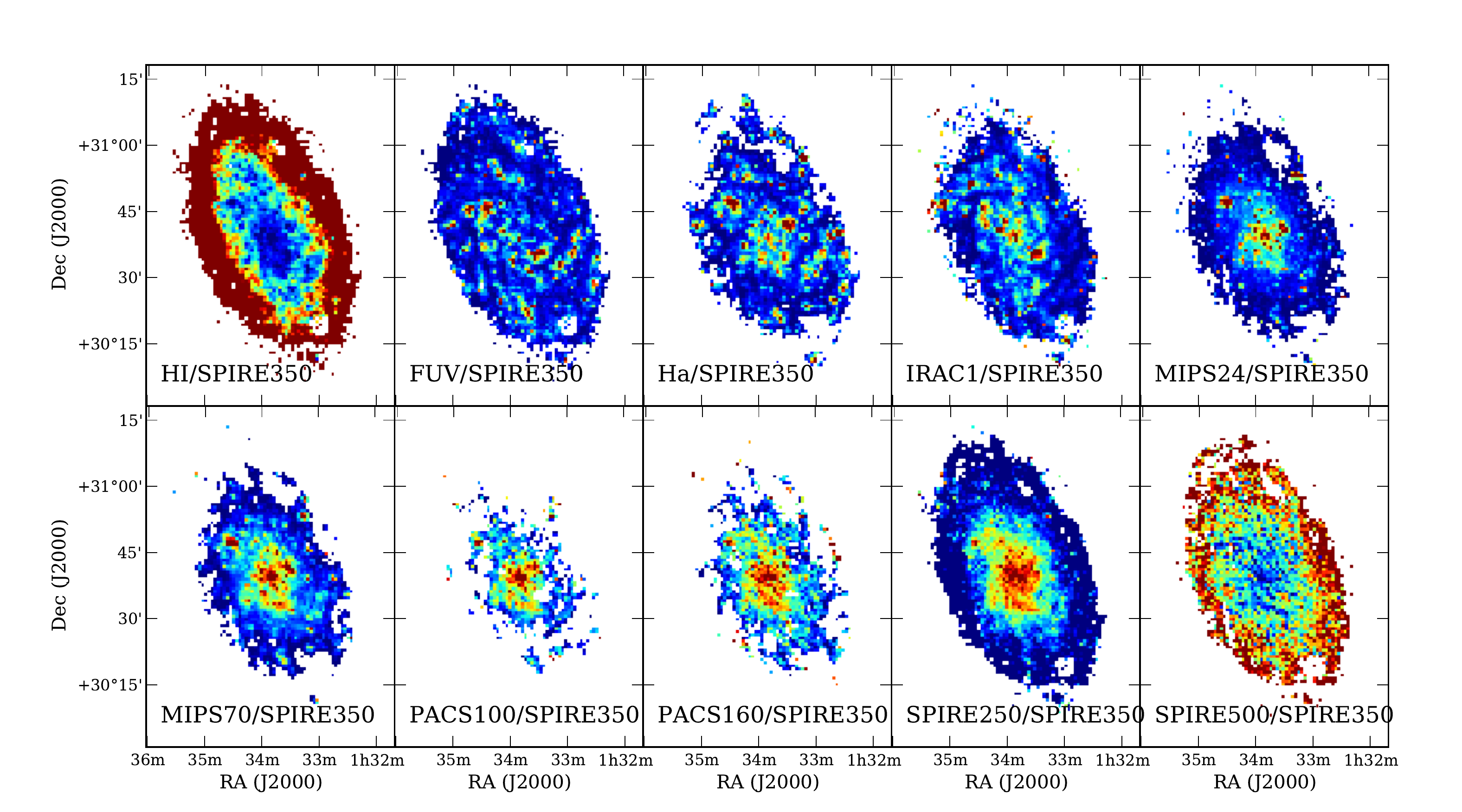}
\includegraphics[width=\textwidth]{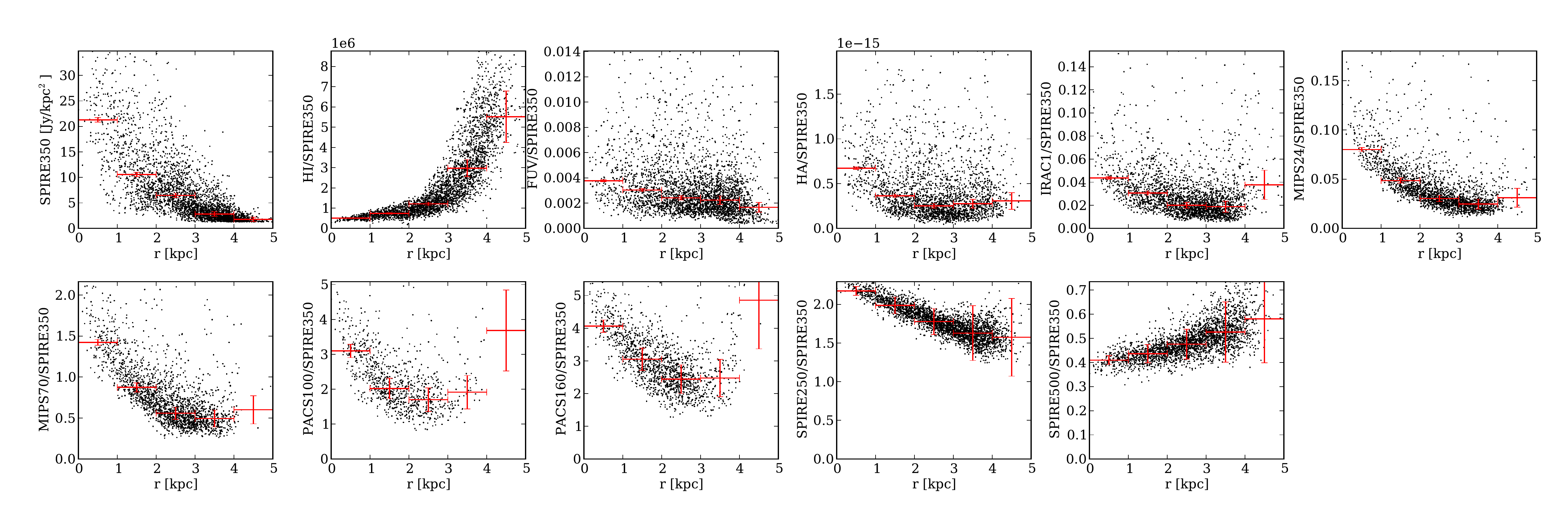}
\includegraphics[width=\textwidth]{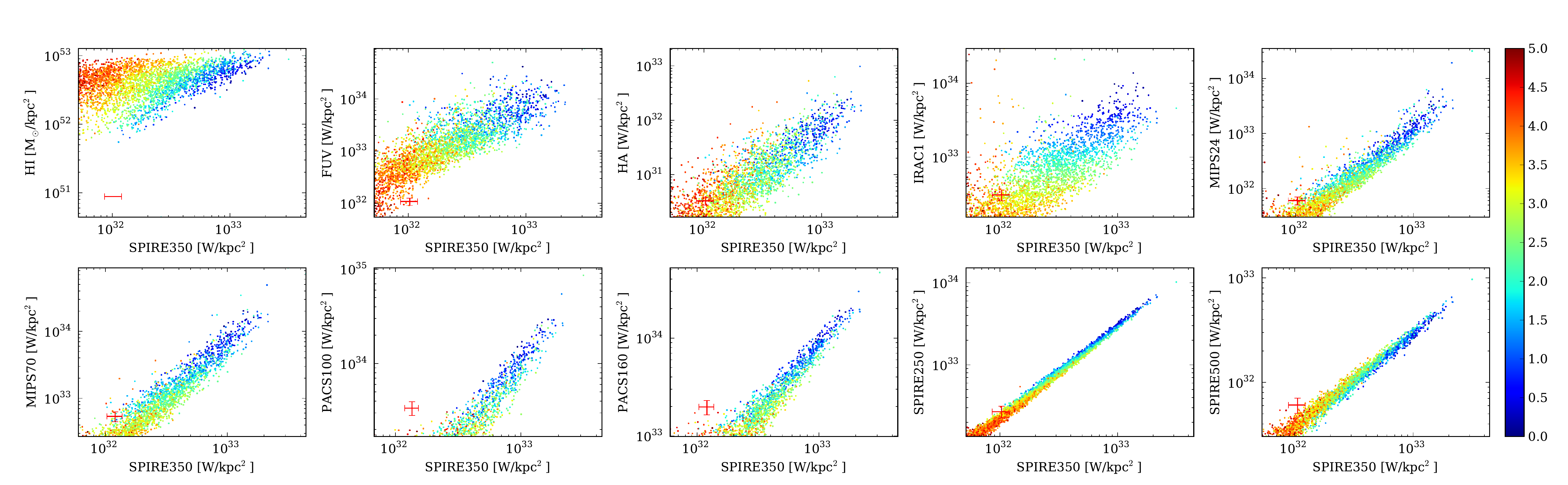}
\caption{Same as in Fig.~\ref{fig:MIPS24} but for the 350~$\mu$m band surface brightness.\label{fig:SPIRE350}}
\end{figure*}
\end{center}

The cold dust is traced by the emission at 250~$\mu$m, 350~$\mu$m, and 500~$\mu$m as we can see in Fig.~\ref{fig:SPIRE350}.

Compared to the atomic neutral gas, the cold dust emission decreases when we go outward, possibly due to the progressively lower radiation field intensity. However, at a given radius the dust emission and the HI mass are well correlated. The emission from the evolved stellar populations seems to diminish faster than the one of the cold dust when going outward. If a slight radial gradient is visible, some spiral structure is also visible probably indicating the larger density of cold dust in spiral arms. A few regions that present little cold dust emission compared to the stellar populations are actually located in HI holes in the galaxy. The warm dust being enhanced in the central regions shows that the relative importance of the cold dust increases with radius. Finally, the cold dust colors seem to be driven by different factors. Indeed, looking at 250/350 and 250/500 we see both a radial gradient and a spiral structure. However, for the 350/500 ratio the radial gradient is much shallower and the spiral structure fainter than for 250/350 and 250/500. The correlation between 250~$\mu$m and 350~$\mu$m is remarkably tight compared to other correlations, with a slight dependence on the distance at a given surface brightness, translating the radial gradient. It is even tighter than the correlation between the 350~$\mu$m and the 500~$\mu$m bands. This can also be seen examining the radial trends. We will discuss this aspect in Sec.~\ref{sssec:youngnold}.

\section{Discussion}
\label{sec:discussion}

\subsection{Estimation of the TIR brightness}
\label{ssec:TIR}
\begin{center}
\begin{figure}[!htbp]
\includegraphics[width=\columnwidth]{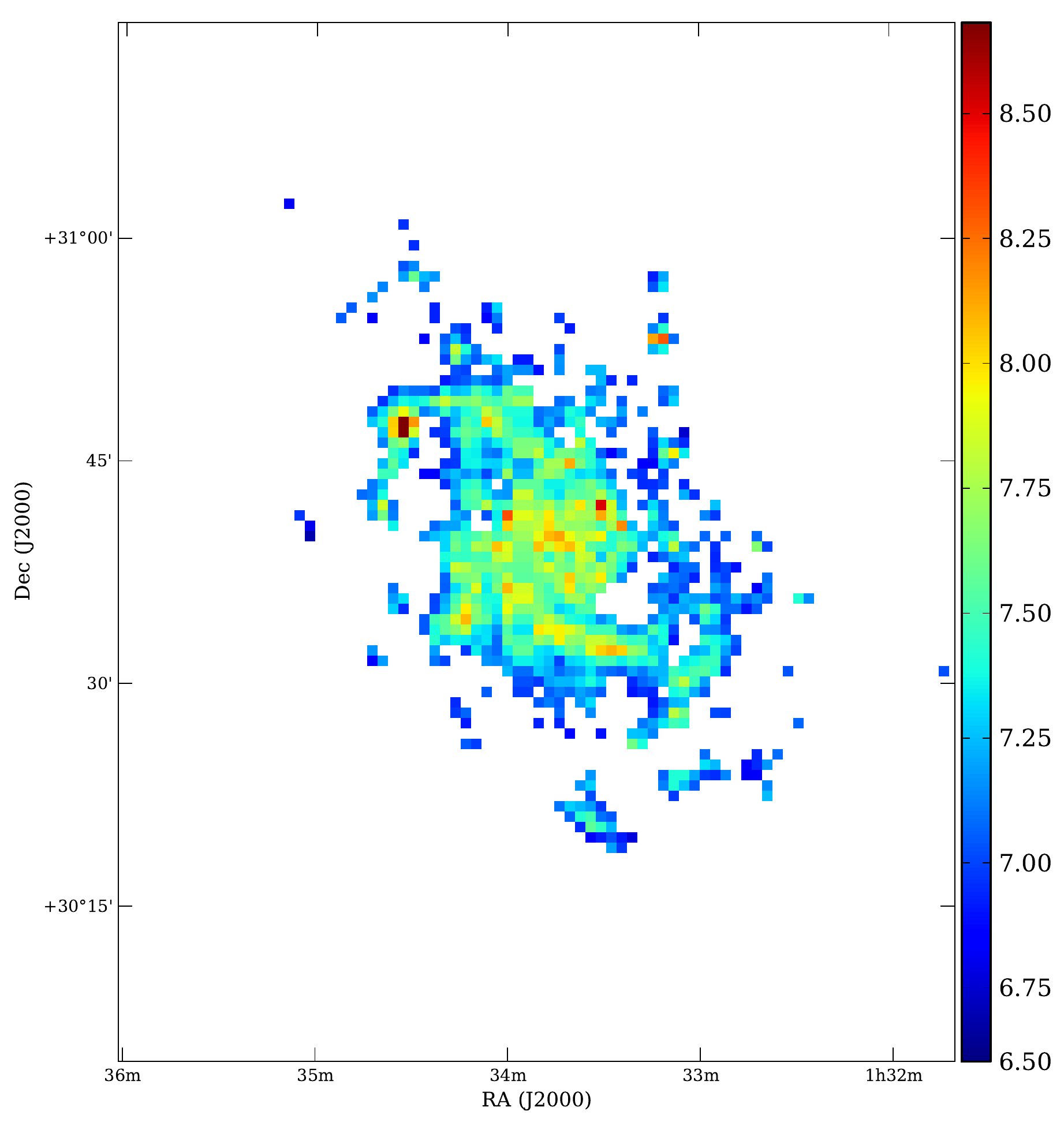}
\caption{Map of the TIR emission. The colorbar on the right indicates the decimal log of the brightness in units of solar luminosity per kpc$^2$. In each pixel, we fit a \cite{draine2007a} model leaving the minimum and maximum starlight intensities $\mathrm{U_{min}}$ and $\mathrm{U_{max}}$ as free parameters. Selecting the best fit we integrate the resulting SED from 1~$\mu$m to 1~mm to obtain the TIR brightness. Only pixels detected at a 3--$\sigma$ level in all bands from 8.0~$\mu$m to 500~$\mu$m are selected here.\label{fig:tir}}
\end{figure}
\end{center}

The TIR brightness from 1~$\mu$m to 1~mm is a proxy for the amount of star formation. With the advent of new instruments such as ALMA, estimating the TIR from just one or two bands is important, in particular for the study of high redshift galaxies. \cite{boquien2010a} published relations to estimate the TIR from Spitzer bands, using regions within galaxies as well as entire galaxies. Here we extend this study to M33 using Herschel bands from 100~$\mu$m to 500~$\mu$m. To estimate the TIR brightness, we fit the \cite{draine2007a} models for each pixel using the stellar subtracted 8~$\mu$m, 24~$\mu$m, 70~$\mu$m, 100~$\mu$m, 160~$\mu$m, 250~$\mu$m, 350~$\mu$m, and 500~$\mu$m bands through a $\chi^2$ minimization. To determine the fluxes from the models and to account for the color corrections, we convolve the SED (spectral energy distribution) with the filter bandpasses. Using the best fit, we then integrate the SED of the dust from 1~$\mu$m to 1~mm. More details about results obtained through the model fittings are published in Rosolowsky et al. (in preparation). In Fig.~\ref{fig:tir}, we present the map of the computed TIR.

To estimate the TIR brightness from just one or several IR bands we compute the best fit of the form:
\begin{equation}
\label{eqn:TIR}
\log S_{TIR}=\sum_{i=1}^n a_i \log S_i+b,
\end{equation}
where $S_{TIR}$ is the TIR brightness, $i$ is the index of the IR band, $n$ is the number of bands we use to determine $S_{TIR}$, and $S_i$, defined as $\nu S_\nu$, is the brightness in the band $i$ in units of W~kpc$^{-2}$. The fit in log--log space is necessary to take into account the non--linearities of the relation between the TIR and the monochromatic emission in some bands. We present the coefficient for the best fit for every combination of Spitzer and Herschel bands from 8.0~$\mu$m to 500~$\mu$m in Table~\ref{tab:fit-tir}.

\begin{center}
\begin{figure}[!htbp]
\includegraphics[width=\columnwidth]{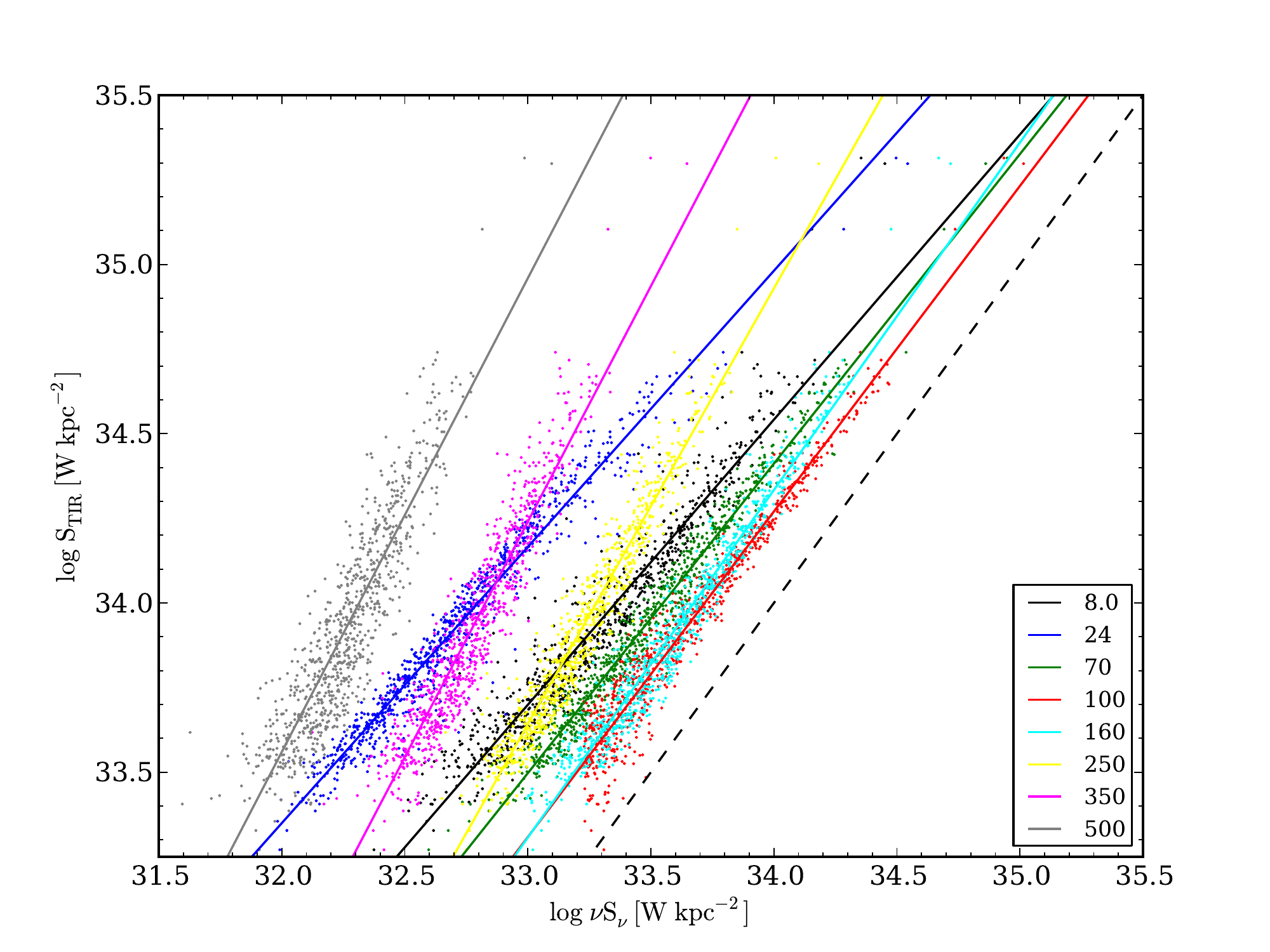}
\caption{Estimate of the TIR brightness from each one of the 8.0, 24, 70, 100, 160, 250, 350, and 500~$\mu$m bands. Each point represents a single pixel in the registered images at the SPIRE 500~$\mu$m resolution. The best fit for each band is shown by a solid line of the same color, the numerical parameters are listed in Tab.~\ref{tab:fit-tir}. The black dashed line represents the 1--to--1 relation, for which a band would account for the totality of the TIR brightness.\label{fig:correlation}}
\end{figure}
\end{center}

In Fig.~\ref{fig:correlation}, we show the fits to evaluate the TIR brightness from each individual IR band.

We see that the 100~$\mu$m and 160~$\mu$m bands are good estimators of the TIR brightness. At lower brightness, the 160~$\mu$m band tends to dominate, with the share of the 250~$\mu$m band increasing. The 8~$\mu$m and 24~$\mu$m bands are sub--linear estimators of the TIR brightness, which is consistent with what \cite{bavouzet2008a,boquien2010a} found both for entire galaxies and regions in galaxies. However it has to be noted that \cite{zhu2008a} found the 8~$\mu$m band to be a nearly linear estimator. The warm dust is close to being a linear estimator of the TIR whereas the cold dust presents ever stronger non--linearities at longer wavelengths owing to the fact that the warm dust drives the TIR brightness. Our coefficients are comparable to the ones of \cite{boquien2010a}, when using only 1 band, showing that the properties of M33 do not differ strongly from the typical properties of their sample which was based on SINGS \citep[Spitzer Infrared Nearby Galaxies Survey,][]{kennicutt2003a}, LVL \cite[Local Volume Legacy Survey,][]{lee2008a} and the \cite{engelbracht2008a} star--forming and starburst galaxies sample. Also, Mookerjea et al. (2011, submitted) find very similar results estimating the TIR from our relation for the 160~$\mu$m band and using their own method based on a combination of gray bodies.

\begin{center}
\begin{figure}[!htbp]
\includegraphics[width=\columnwidth]{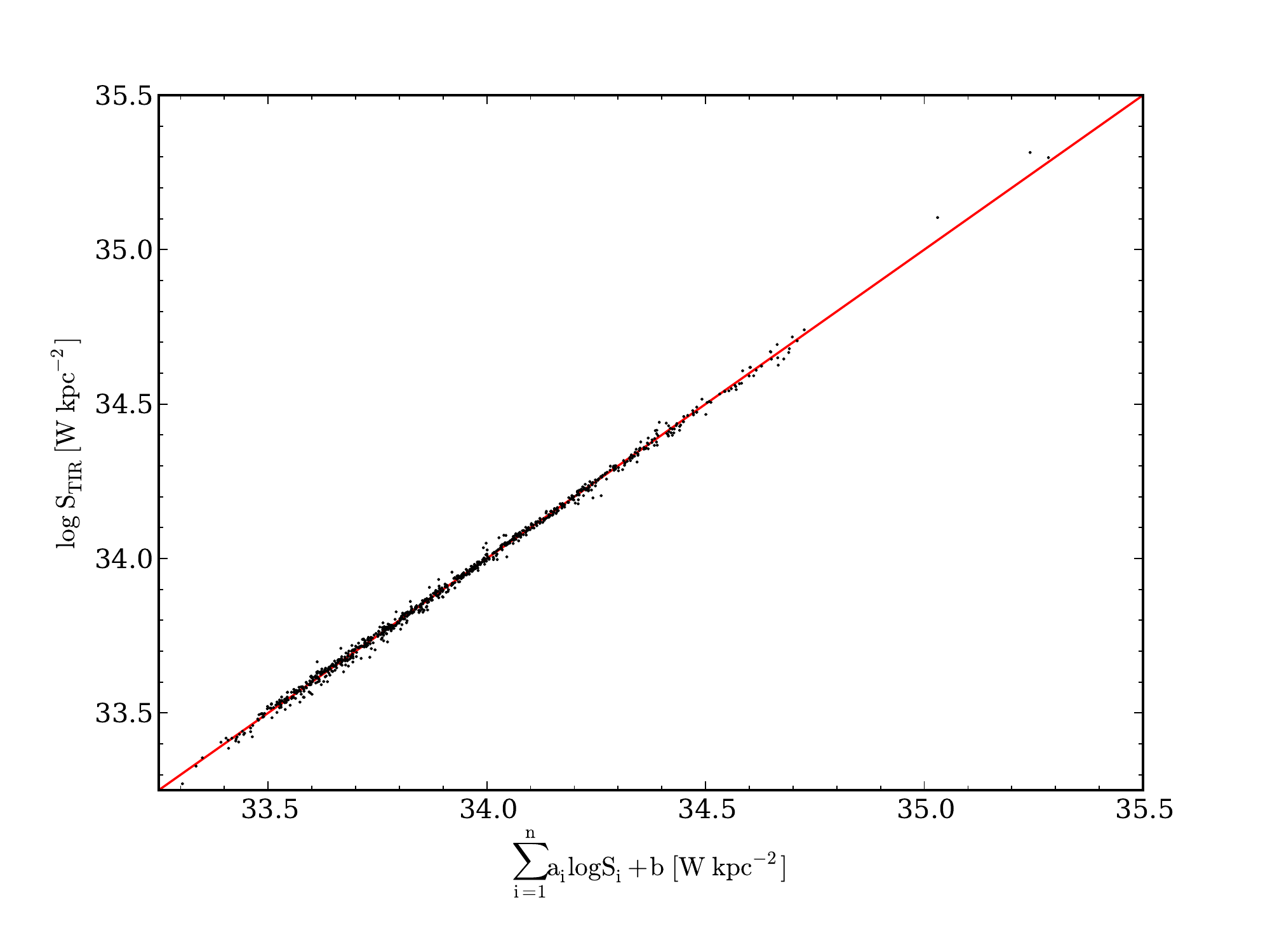}
\caption{Estimate of the TIR brightness from the combination of the 8.0, 24, 70, 100, 160, 250, 350, and 500~$\mu$m bands. The red line represents the 1:1 relation between the TIR estimated from each IR band and the one estimated from the fit of the SED with a \cite{draine2007a} model. The parameters of the fit are listed in Tab.~\ref{tab:fit-tir}. Note that the range is similar to that of Fig.~\ref{fig:correlation}. A total of 850 pixels are taken in account in this fit.\label{fig:correlation-all}}
\end{figure}
\end{center}

In Fig.~\ref{fig:correlation-all} we plot the best fit using all IR bands available.

We see that the fit using all bands is very tight with few points deviating from the expected TIR brightness with a scatter around the best fit of 0.012~dex.

\subsection{Star formation rate estimators}
\label{ssec:SFR}

Recently, \cite{boquien2010b} and \cite{verley2010b} showed that using individual star--forming regions, the emission in the 100~$\mu$m to 250~$\mu$m bands is a nearly linear tracer of the SFR. However, star--forming regions in more distant galaxies cannot necessarily be disentangled from one another even if Herschel resolves the galaxy as a whole. As a consequence, we examine here what is the best combination of Spitzer and Herschel bands to estimate the SFR.

Star--formation is typically traced either by the photospheric emission of young, massive stars (UV), recombination lines of the gas that they ionize (H$\alpha$), the dust they heat (mid-- and far--IR), or the radio continuum from free--free thermal emission in HII regions and synchrotron emission in supernovae remnants. They have different properties regarding their sensitivities to the star formation history, the extinction, the metallicity, etc. The use of the IR as a SFR estimator is affected by two mains factors. First, except for the dustiest galaxies, a fraction of the radiation emitted by young stars escapes from the surrounding gas clouds without heating the dust. This may result in an underestimate of the SFR in star forming regions. This effect can be estimated combining the IR data with bands where this radiation emerges, such as UV or H$\alpha$ \citep{calzetti2007a,kennicutt2007a,leroy2008a,bigiel2008a,kennicutt2009a}. The second factor is that the radiation from older stellar populations can heat the dust. This is particularly prevalent in early--type galaxies \citep{sauvage1992a}. As a consequence, using the IR may result in an overestimate of the SFR. However, this problem should be less severe in the case of strongly star--forming galaxies and late--type galaxies, such as M33, as there are proportionally more young stellar populations compared to older ones.

\cite{calzetti2007a} showed that the current star formation is accurately traced in HII regions by a combination of the H$\alpha$ with the 24~$\mu$m emission:\begin{equation}
\mathrm{SFR=5.35\times10^{-35}\left[L\left(H\alpha\right)+0.031\times L\left(24\right)\right]},
\end{equation} where L is in W (with L(24) defined as $\mathrm{\nu L_\nu}$ at 24~$\mu$m) and $5.35\times10^{-35}$ the conversion factor to the SFR in units of M$_\sun$~yr$^{-1}$, assuming a \cite{kroupa2001a} initial mass function. We assume that all the emission is linked to star formation. This assumption is mostly valid in bright regions. For fainter regions it depends on the fraction of the emission due to other processes as we have exposed in the introduction. To estimate the SFR from any combination of Spitzer and Herschel bands, we follow the same procedure as presented in Sec.~\ref{ssec:TIR}:

\begin{equation}
\log SFR=\sum_{i=1}^n a_i \log S_i+b, 
\end{equation}
where SFR is in M$_\odot$~yr$^{-1}$~kpc$^{-2}$ and the right--hand side of the relation is defined as in Eq.~\ref{eqn:TIR}. The parameters of the best fits are presented in Tab.~\ref{tab:fit-sfr}.

\begin{center}
\begin{figure}[!htbp]
\includegraphics[width=\columnwidth]{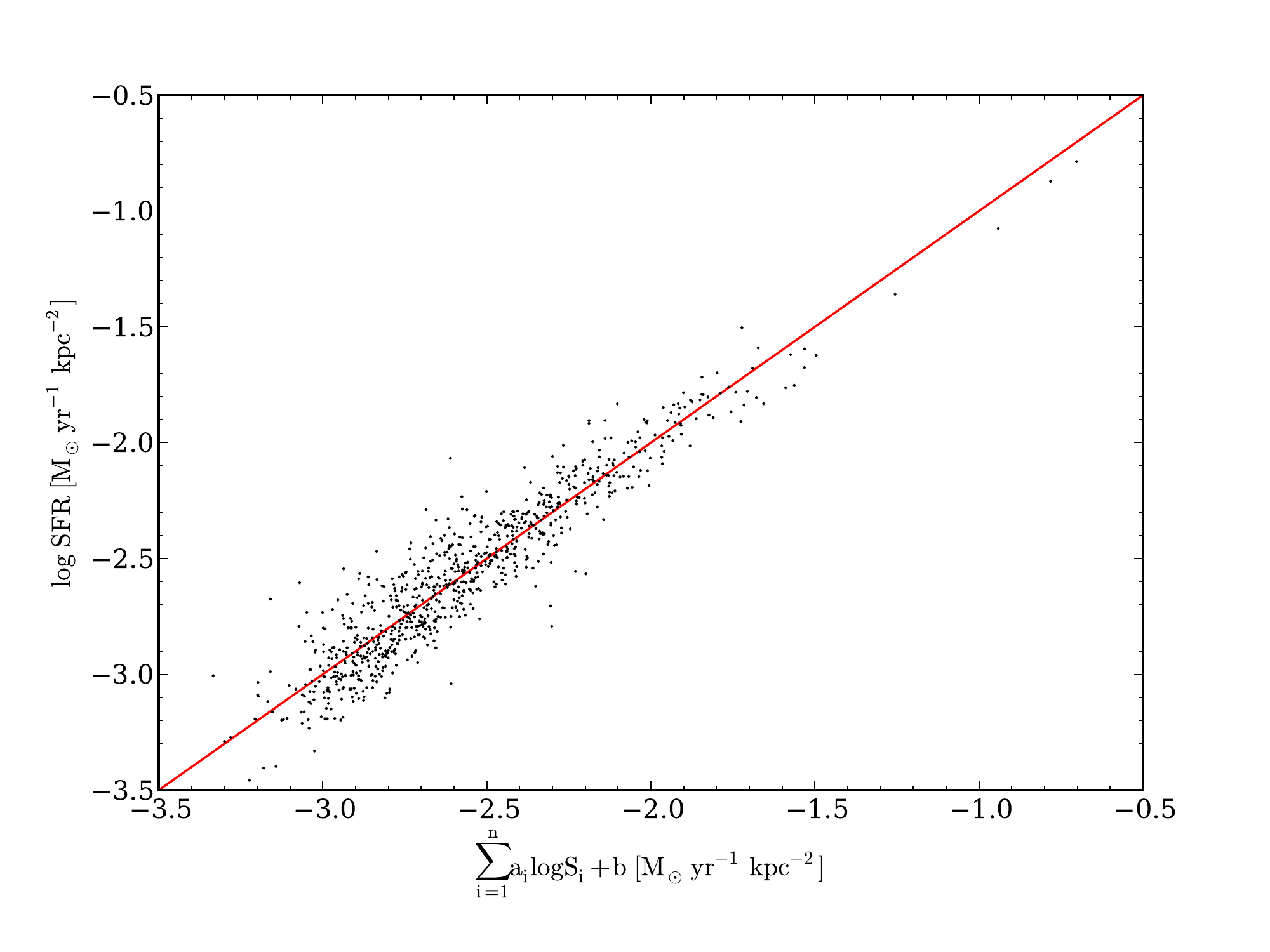}
\caption{Estimate of the SFR from the combination of the 8.0, 24, 70, 100, 160, 250, 350 and 500~$\mu$m bands. The red line represents the 1:1 relation between the SFR estimated from each IR band and the one estimated from the combination of the H$\alpha$ and 24~$\mu$m images. The parameters of the fit are listed in Tab.~\ref{tab:fit-sfr}.\label{fig:correlation-sfr-all}}
\end{figure}
\end{center}

First of all, when we estimate the SFR from only one band, we find that both the 8.0~$\mu$m and the 24~$\mu$m emission are sub--linear tracers of the SFR. If the 24~$\mu$m slope is consistent with the values provided by \cite{calzetti2010a}, the case of the 8~$\mu$m is surprising. \cite{calzetti2007a} for instance found a super--linear relation. M33 having a sub--solar metallicity, may explain why we find this difference compared to the higher metallicity sample of \cite{calzetti2007a}. Another possible explanation is that \cite{calzetti2007a} selected HII regions whereas we select all regions above a given SNR, therefore also including quiescent regions. This has 2 main consequences. First the SFR yielded by the combination of the H$\alpha$ and 24~$\mu$m emission is not as reliable in faint regions as in HII regions as mentioned earlier. Then, it is well known that PAHs get destroyed more easily in bright regions where the radiation field is intense, which is not the case in fainter regions. We will see in Sec.~\ref{sssec:non-linearities} that if we eliminate the faintest regions the slope quickly becomes super--linear as is expected by previous results. When looking at the warm dust from 70~$\mu$m to 160~$\mu$m, all bands are close to being linear tracers. The 70~$\mu$m band is a nearly linear estimator of the SFR which corroborates the finding of \cite{li2010a} on individual star--forming regions in galaxies with $12+\log O/H>8.4$. Interestingly, \cite{boquien2010b} found that the 100~$\mu$m is a linear tracer and that the 160~$\mu$m was slightly superlinear, based on individual star--forming regions. The cold dust proves to be a significantly non--linear tracer of star--formation. \cite{verley2010b} found that the 250~$\mu$m band was a linear tracer when selecting individual star--forming regions and estimating the SFR in a similar way as we do here. The reason is that \cite{verley2010b} derived the SFR from the luminosity of individual star forming regions. When deriving a relation using the surface brightness they also find a super--linear slope close to what we find here. Finally, when combining all bands, the scatter around the best fit is reduced to 0.128~dex. The best fit using all bands is presented in Fig.~\ref{fig:correlation-sfr-all}.

We have also estimated the SFR using the FUV image corrected for the extinction with the 24~$\mu$m using the calibration presented by \cite{leroy2008a,bigiel2008a}. This yields much shallower slopes that are most likely underestimated. A fit shows that $\mathrm{SFR(FUV+24)\ \propto \ SFR(H\alpha+24)^{0.78}}$. The reason is probably that at such a refined spatial scale, age effects are particularly significant. Such a dependence on the way the SFR is estimated was shown by \cite{bigiel2008a}.

\subsection{Heating sources of the dust populations}
\subsubsection{Radiation field hardness}
Recently, studying a sample of galaxies from the Herschel Reference Survey \citep{boselli2010a} and from the Herschel Virgo Cluster Survey \citep{davies2010a} along with M81 \citep{bendo2010a} and M82 \citep{roussel2010a}, \cite{boselli2010b} showed that the warm dust temperature, measured by the 60/100 ratio, increases with the birthrate parameter $b$, defined as the current SFR divided by the mean SFR over the lifetime of the galaxy, whereas the cold dust temperature, measured by the 350/500 ratio, decreases. The birthrate parameter is relevant to examine the heating sources of the dust as it traces the hardness of the radiation field. To determine the birthrate parameter, we estimate the SFR similarly as before, combining the 24~$\mu$m with the H$\alpha$ emission \citep{calzetti2007a}, and with the FUV one \citep{leroy2008a,bigiel2008a}. Also, to determine the stellar mass, we use the 3.6~$\mu$m data we presented earlier. To convert the 3.6~$\mu$m luminosity to the stellar mass we use the relation presented by \cite{oliver2010a} for Scd galaxies: $\mathrm{M_\star=35.3\times\nu L_\nu\left(3.6\right)}$, with $\mathrm{M_\star}$ the stellar mass in $\mathrm{M_\odot}$ and $\mathrm{\nu L_\nu}$ the 3.6~$\mu$m luminosity in $\mathrm{L_\odot}$. We assume the galaxy has been forming stars over the last $13\times10^9$~yr.

\begin{center}
\begin{figure}[!htbp]
\includegraphics[width=\columnwidth]{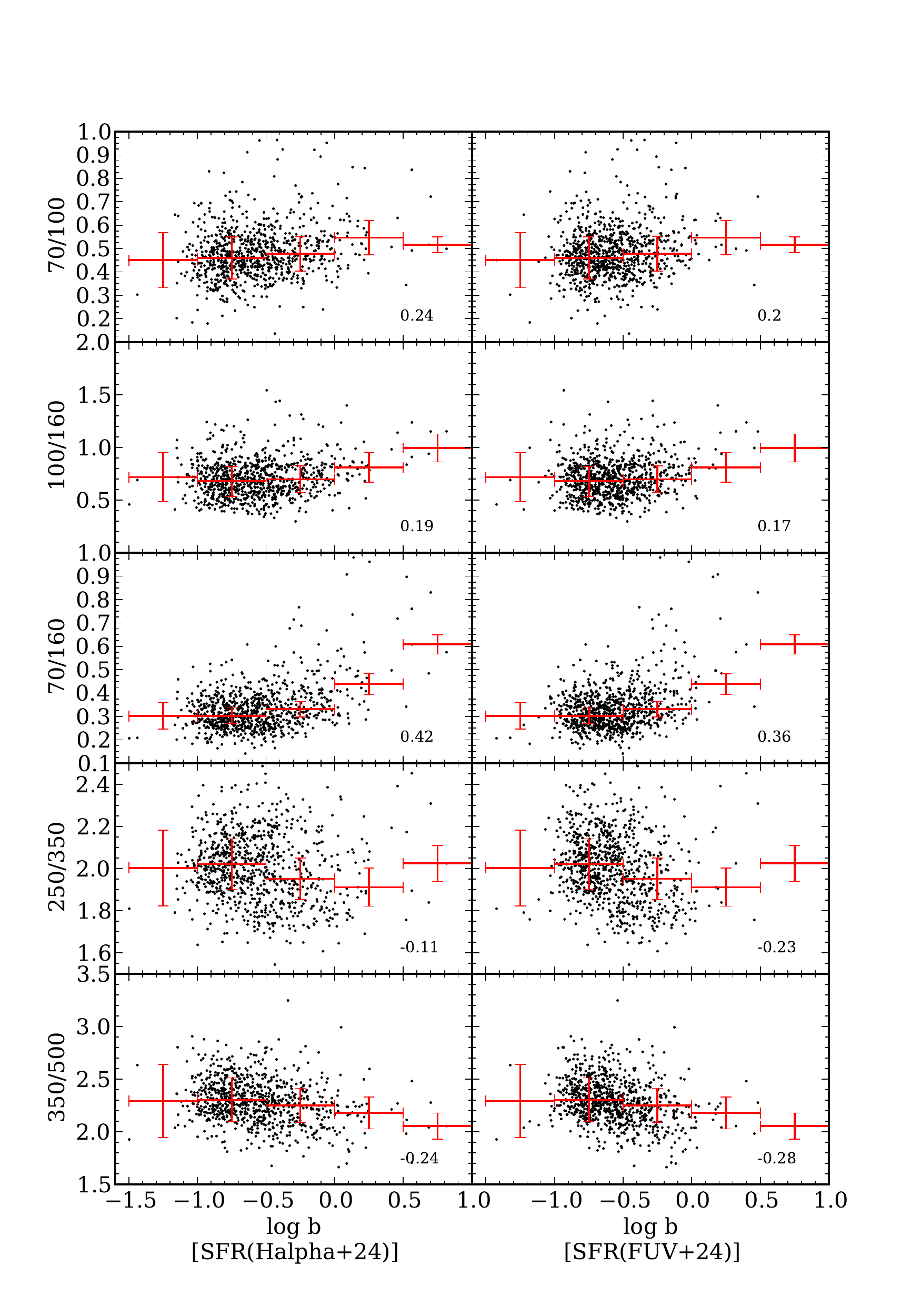}
\caption{Colors of the different dust components as a function of the birthrate parameter.\label{fig:b-trends}}
\end{figure}
\end{center}

Various dust color trends with the birthrate parameter are presented in Fig.~\ref{fig:b-trends}. We find the same trend as \cite{boselli2010b}, namely the warm dust color, defined as the ratio of 2 IR bands, is correlated with the birthrate parameter whereas the cold dust color tends to be anti--correlated. In both cases the correlation is much weaker though. We do not observe any significant difference depending on whether the SFR is derived from the H$\alpha$ emission or from the FUV emission. For the rest of the analysis, we concentrate on the former case. Upon closer examination, it appears that for the warm dust, the trend is driven overwhelmingly by regions that have $\log b>-0.5$. If we select only pixels under this threshold, for the 70/160 color the Pearson correlation coefficient falls to 0.03 which means it is uncorrelated, and increases to 0.50 for those above it which means it is marginally correlated. This can be easily understood as regions with a particularly high SFR will tend to have a large birthrate parameter, even if the underlying stellar mass is important. A high SFR yields bluer dust colors as we will see in the following section. When the SFR is lower, the trend disappears. The anti--correlation of the cold dust with the birthrate parameter is also entirely due to regions with a high birthrate parameter. Setting the same threshold as earlier, for the 350/500 color, the Pearson correlation coefficient reaches to $-0.06$ for pixels under the threshold and $-0.17$ for those above it. In any case these effects are at most extremely weak. Nevertheless if they happen to be real, regions with a high SFR tend on average to be more dusty. As a small quantity of dust absorbs the local radiation (both from star forming regions and from evolved populations), it shields the rest of the dust from this radiation. and as a consequence the rest of the dust tends to get colder. This effect is well known in (U)LIRGs which have cool dust \citep{dunne2001a,calzetti2010a}.

\subsubsection{Young stars or old stars}
\label{sssec:youngnold}
One of the main heating sources of dust is obviously star formation. Young stars are enshrouded in dust clouds and a large fraction of their radiation is absorbed by the dust, which makes IR a star formation tracer as we saw in Sec.~\ref{ssec:SFR}. However, the origin of diffuse dust emission in galaxies is not well understood. This dust could either be heated by the energetic radiation of young stars escaping from star forming regions, which could be absorbed at large distances, or it could also be heated locally by the field stellar population. Using Herschel observations of the early--type galaxy M81, \cite{bendo2010a} found that the cold dust is heated by old stars whereas the warm dust is heated by younger stars and the presence of an active nucleus.

\begin{center}
\begin{figure*}[!htbp]
\includegraphics[width=\textwidth]{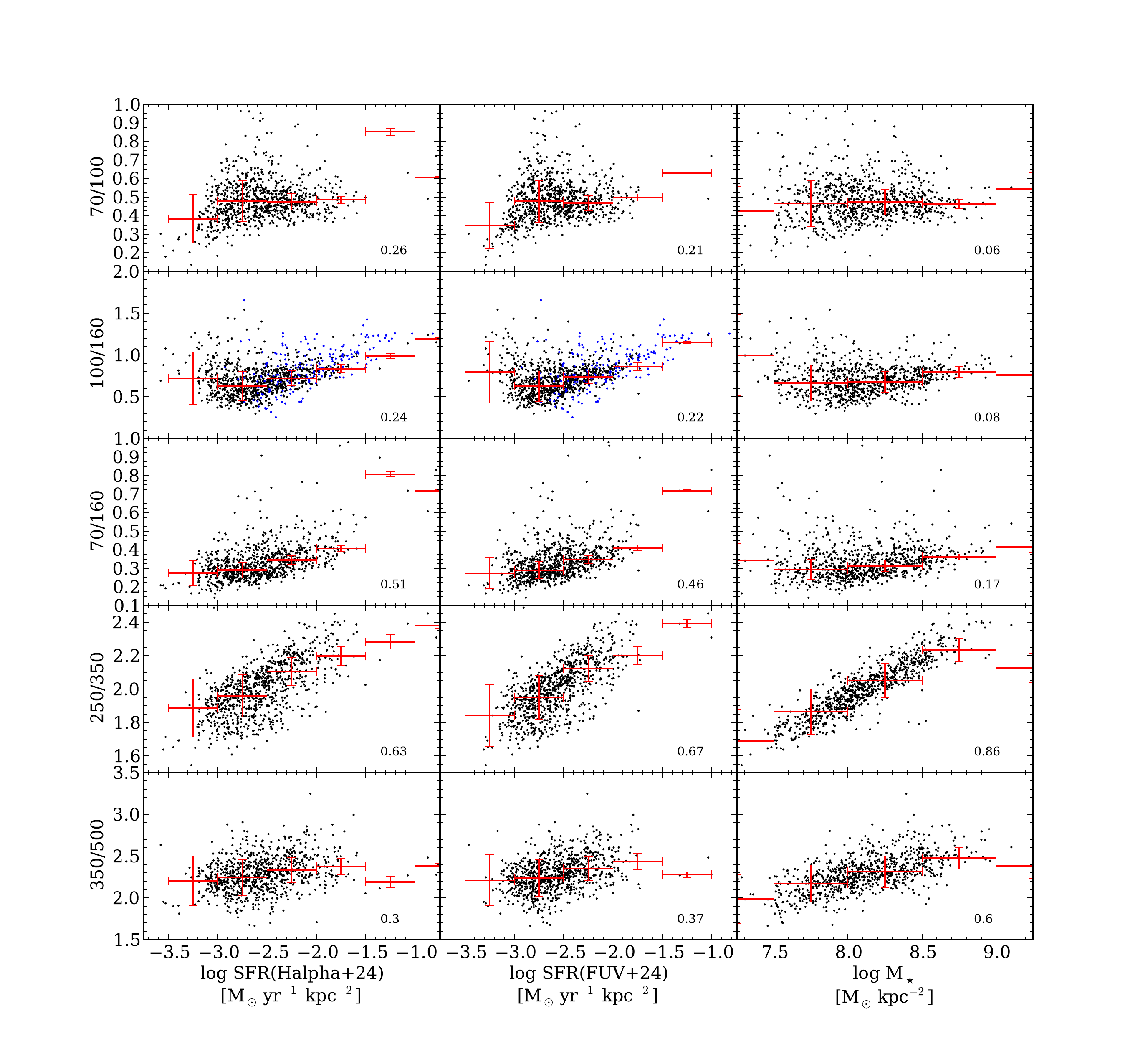}
\caption{Temperature, as measured by the color of the warm and cold dust populations versus the SFR (derived from both H$\alpha$ and FUV which are combined with the emission at 24~$\mu$m) and the local stellar mass per unit area. The points in blue correspond to integrated HII regions of \cite{boquien2010b} for which the local background in M33 has been removed, subtracting in effect the contribution from the local diffuse emission. The error bars in red indicate the median uncertainty in bins that are respectively $10^{0.5}$~M$_\sun$~yr$^{-1}$~kpc$^{-2}$ and $10^{0.5}$~M$_\sun$~kpc$^{-2}$ wide. The Pearson correlation coefficient is indicated in the bottom--right corner of each plot.\label{fig:heating-source}}
\end{figure*}
\end{center}

To test for the heating sources, we plot in Fig.~\ref{fig:heating-source} the color of the warm and cold dust components versus both the SFR and the local stellar mass. 

Unsurprisingly, the colors show some correlation with both the SFR and the stellar mass per unit area. This is expected as those two variables are not truly independent, the inner regions that have more evolved stars also have more star formation. As the results do not depend significantly on the band from which the SFR is derived, we consider here only the case where the SFR is derived from H$\alpha$ and 24~$\mu$m. Regarding the warm dust, the correlation coefficient is higher for the SFR compared to the one for stellar mass, whereas it is the contrary in the case of the cold dust. For the warm dust, we can distinguish 2 different regimes in the 100/160 ratio, with a cut--off around $10^{-2.75}$~M$_\odot$~yr$^{-1}$~kpc$^{-2}$. If we select only pixels under this threshold, the Pearson correlation coefficient drops to $-0.29$. Conversely for pixels above this threshold, the correlation is tighter with the correlation coefficient up to $0.49$. Those 2 regimes are also seen when using the longer baseline ratio 70/160. Under the threshold, only little trend is seen (0.19) whereas it is clearer above it (0.47). This hints that at lower SFR, other processes have an influence on the temperature of the warm dust. The most obvious explanation would be that at the lowest SFR the radiation field is actually dominated by the evolved stellar populations whereas above this threshold, star formation is the dominant mechanism. Indeed, even if young stars represent only a small fraction of the total stellar mass, they easily dominate the UV radiation field. If we consider the past 10~Myr, a relevant timescale for H$\alpha$ and 24~$\mu$m in star forming regions, the mass of young stars formed is $10^{4.25}$~M$_\odot$~kpc$^{-2}$ if we assume a SFR of $10^{-2.75}$~M$_\odot$~yr$^{-1}$~kpc$^{-2}$. To dominate the UV radiation field, a young population typically needs to represent at least $10^{-3}$ to $10^{-4}$ of the total stellar mass. This corresponds to masses from $10^{7.25}$ to $10^{8.25}$~M$_\odot$~kpc$^{-2}$. The majority of selected pixels have stellar masses in this range, giving weight to this explanation. Interestingly, when examining the SFR map, setting a threshold at $10^{-2.75}$~M$_\odot$~yr$^{-1}$~kpc$^{-2}$, it roughly delineates the spiral arms and the inner parts of the galaxy, typically at a inclination--corrected distance smaller than 1.2~kpc from the center of M33. When looking at the 100/160 ratio of individual star forming regions with the local background subtracted \citep{boquien2010b}, the overall Pearson correlation coefficient is higher (0.44). This is another hint at the influence of the evolved stellar populations on the warm dust temperature. In any case, these results show that, at low brightness levels, the warm dust emission cannot be simply a result of local heating by young stellar populations, but other factors, such as heating from evolved populations, circumstellar dust from post--AGB stars whose SED peaks in that range \citep[][for example]{hoogzaad2002a,verley2007a}, as well as radiation escaping from star--forming regions, need to be accounted for.

The correlation of the 250/350 ratio with the stellar mass is particularly clear, showing without ambiguity the importance of the evolved stellar populations driving the temperature of the cold dust. Looking at the 350/500 ratio, while the trend is still visible, the correlation coefficient is weaker. The slightly larger scatter compared to the uncertainties may be a weak hint that another process is at play in the 500~$\mu$m band, though the evidence is extremely tenuous. Among possible explanations would be, a pollution by very cold dust, the variation of the emissivity of the cold dust at longer wavelength or changes in the physical properties of the dust \citep{israel2010a,bot2010a}. This question will be addressed in upcoming HERM33ES papers using PACS and SPIRE data complemented by MAMBO2 1.2~mm and LABOCA 870~$\mu$m data (Quintana--Lacaci et al. in prep., Anderl et al. in prep.).

Looking simultaneously at the warm and cold dust components can also provide us with information regarding dust heating sources. For instance as we see in Fig.~\ref{fig:PACS160}, the relative scatter of PACS/SPIRE bands increases with radius. To separate the variation of the scatter that is due to intrinsic measurement uncertainties from the one that is due to physical phenomena in M33, we have quadratically subtracted the observed scatter of the ratio of the 2 bands from the one determined through the noise measurements on the images. In some cases the increase of the scatter can be entirely attributed to measurement uncertainties. However in other cases the scatter is higher than what is expected from measurement uncertainties only. As an example in the case of 100/250, in bins 1~kpc wide the final relative scatter to the median goes from $0.18$ for regions under 1~kpc up to $0.47$ for regions between 3 and 4~kpc. Similarly for 160/250 it goes from $0.12$ to $0.40$. Such an increase in the scatter may be due to a change of the dust heating sources: in the inner regions with large quantity of young and old stars creates a rather uniform radiation field. Conversely in the outer regions, the radiation field is more dominated by evolved stellar populations with the occasional OB stars locally boosting the radiation field. This hint is to be taken with great caution as it is heavily dependent on the accuracy of uncertainty measurements which may be biased as explained in Sec.~\ref{ssec:flux-measurement}.

\subsection{Limitations and their impact on the results}
\subsubsection{Calibration uncertainties\label{sssec:calib-error}}

Bolometers are relative rather than absolute detectors, which makes the calibration of the images delicate. While this should not affect our results that are based on trends, it is more problematic for quantitative results in the determination of the TIR and the SFR for instance.  The most notable problem lies with the PACS data. When comparing PACS~160~$\mu$m with MIPS~160~$\mu$m images, the fluxes are increasingly discrepant as they are higher, with a non--linear slope of $\sim1.3$ in log space leading to an emission ``excess'' in PACS maps (or an emission ``deficit'' in the MIPS ones) for brighter regions. This overestimate primarily affects the SED fits with the \cite{draine2007a} models. Upon examination, it appears that the distribution of the difference between the observed flux and the best model is non--gaussian and asymmetric when using the PACS~160~$\mu$m band, whereas no clear trend is seen either way in the case of SPIRE~250~$\mu$m for instance. This could affect the estimate of the TIR brightness and of the SFR as the PACS bands dominate the energy budget in the IR.

\subsubsection{Shallow depth of PACS observations}

The PACS observations are relatively shallow compared to the SPIRE maps which are much closer to the confusion limit. The main consequence is that the diffuse emission that is particularly prevalent in the MIPS 160 map is hardly distinguished in PACS 160~$\mu$m. As we require a SNR of at least 3 in the final convolved, registered maps, much fewer pixels reach that threshold in external regions for PACS bands compared to SPIRE bands, which limits our ability to study faint regions.

\subsubsection{Non--linearities in the estimates of the TIR and the SFR}
\label{sssec:non-linearities}
\begin{center}
\begin{figure}[!htbp]
\includegraphics[width=\columnwidth]{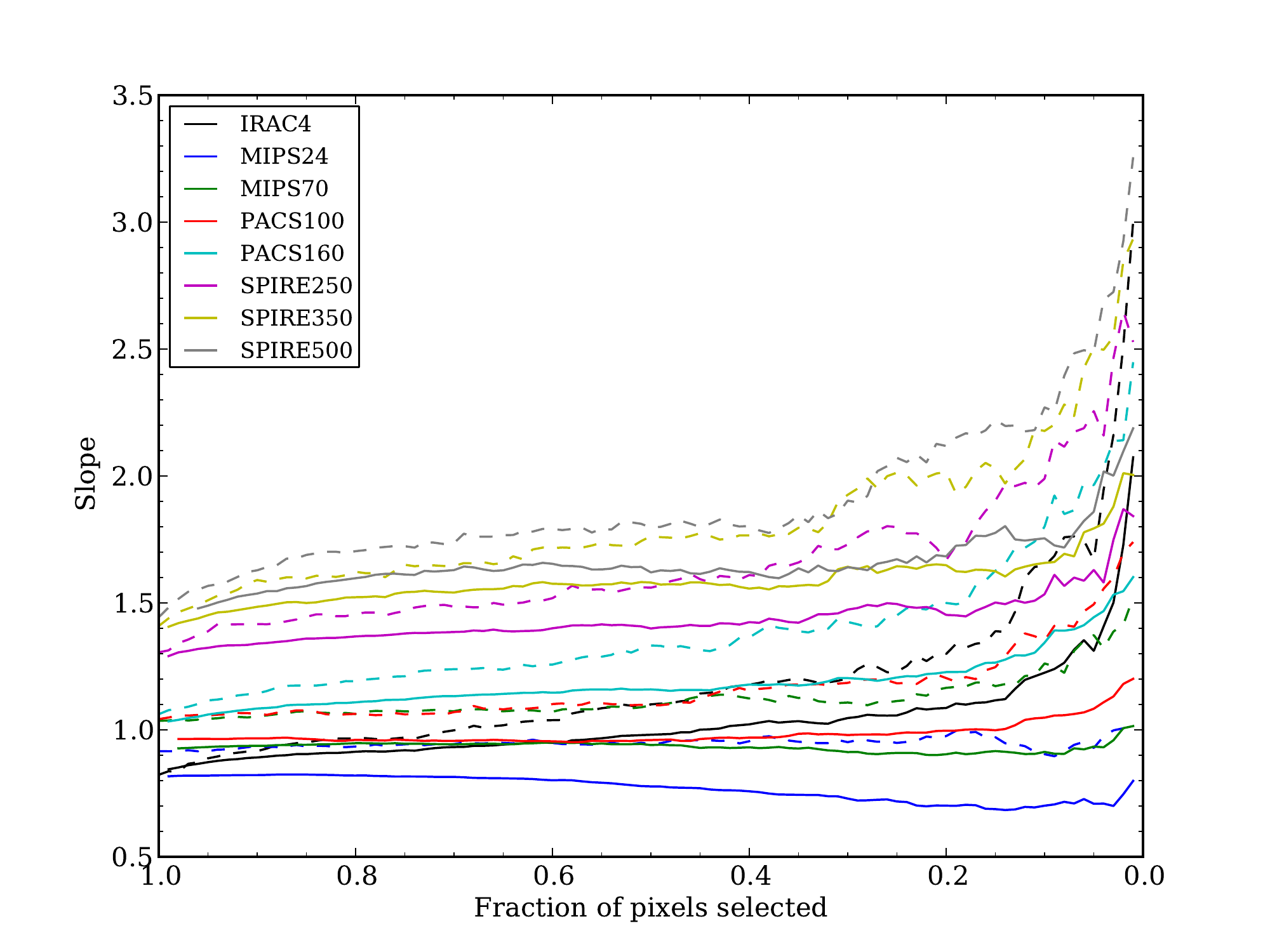}
\caption{Change of the slope when selecting only the brightest pixels in a given band. On the left side of the plot we select all pixels that are detected with SNR$>$3 (850 pixels). We then increase the flux threshold selecting fewer and fewer pixels, eliminating the faintest ones. The solid lines represent the slope is determined the TIR and the dashed lines represent the slope to determine the SFR brightness. \label{fig:slope-change}}
\end{figure}
\end{center}

We have provided estimates of the TIR and of the SFR using one or several infrared bands. To take into account the non--linearities, we have performed the fits in log space. However, depending on the threshold to select the pixels to estimate the TIR and the SFR the slope varies. Indeed, as the threshold increases, that is selecting only the ever brighter pixels, the slope to estimate the TIR or SFR from individual IR bands increases, becoming increasingly non linear, except for the 24~$\mu$m band which shows unclear trends at higher brightness. In Fig.~\ref{fig:slope-change} we plot the slope as a function of the fraction of the brightest pixels selected to determine both the TIR and the SFR.

\begin{center}
\begin{figure}[!htbp]
\includegraphics[width=\columnwidth]{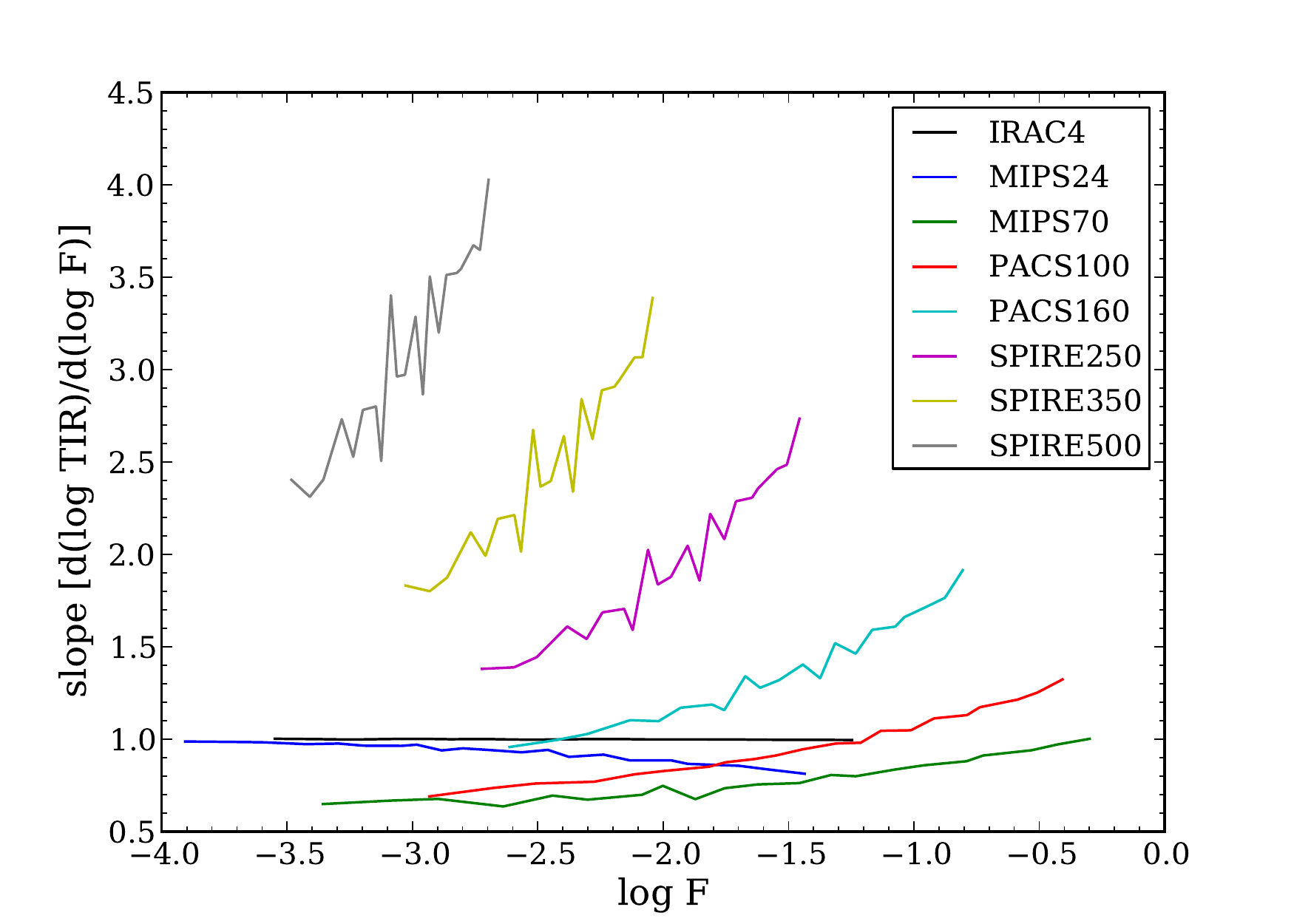}
\caption{Slope $\mathrm{a=d\log TIR/d\log F}$ of the relation to determine the TIR from the brightness $F$ in a given band: $\mathrm{\log TIR = a\log F +b}$. The brightness F is in arbitrary units. The slope is determine from the \cite{draine2007a} model MW3.1\_20. The dust is illuminated by a single radiation field whose starlight intensity varies between U=0.1 and U=25.\label{fig:slope-draine}}
\end{figure}
\end{center}

To understand the origin of this change of slope with increasing brightness, we plot in Fig.~\ref{fig:slope-draine} models from \cite{draine2007a} where we determine the slope as a function of the brightness in one band.

Even though this model only has 1 free parameter and provides the local slope rather than the global slope over a range of brightnesses, we retrieve the same trends described before: the slope from 70~$\mu$m to 500~$\mu$m increases with the brightness, that is selecting only the brightest regions yields a steeper slope. Conversely the slope at 24~$\mu$m decreases. However we can notice that the trend of the 8~$\mu$m band does not properly reproduce the observations. One possible explanation is that the observed trend is largely defined by processes of formation and destruction of PAHs. Indeed the \cite{draine2007a} model used here has a fixed PAH abundance. At higher luminosity the destruction of PAHs can become an important process, which as a result increases the slope. For the SFR, the way it is evaluated can also induce problems as it is not reliable at low surface brightness, which should be dominated by diffuse emission not necessarily linked to a current star formation episode.

While this model is aimed at reproducing trends rather than specific values, the slope for the bands tracing the cold dust seems particularly high (Fig.~\ref{fig:slope-draine}). The model presented here assumes a constant gas mass. In reality, it is well known that actively star forming regions contain proportionally more gas via the Schmidt--Kennicutt law \citep{kennicutt1998b,kennicutt2007a}. It could affect the slope determined from the model, in combination with a variation of the warm--to--cold dust mass ratio as the energy budget of the TIR is dominated by the warm component. Indeed in the model, if the radiation field increases, the energy will be re--radiated by the warm dust with only little change to the emission of the cold dust, which yields a steep slope. However, if we consider an increase of the mass of the gas (and therefore of the dust mass, assuming a constant gas--to--dust mass ratio), the cold dust emission will increase in proportion (assuming its temperature stays constant) and the SFR (or TIR) will increase with a power around 1.4. This explains why the relation between the cold dust and the SFR (or TIR) is closer to 1.4 than what is predicted by the model. Temperature variations in the cold dust component, which are driven by variations in the old stellar population mix, will also play a role. Assuming that the cold dust temperature varies between 15~K and 25~K, the emission will be increased by a factor 2.7 and 3.5 at 500~$\mu$m and 350~$\mu$m respectively, which is much smaller than the dynamical range of the data.


\section{Summary and conclusions}
\label{sec:conclusion}

Using Spitzer and Herschel IR data from 3.6~$\mu$m to 500~$\mu$m, along with GALEX FUV data and HI and H$\alpha$ maps, we have studied the properties and the origin of the warm and cold dust emission in the moderately inclined, local group galaxy M33, at a spatial resolution of 150~pc. To do so, we have studied the evolution of the dust colors both as the position in the galaxy (including the radial distance) and the brightness. We have found the following results:

\begin{enumerate}
 \item The colors of the warm (as seen in the 24~$\mu$m to 160~$\mu$m bands) and cold dust (observed from 250~$\mu$m to 500~$\mu$m) components seem to be predominantly driven by the radiation field intensity.
 \item Combining any set of Spitzer and Herschel bands, we have provided correlations to estimate both the TIR brightness and the SFR, extending the results of \cite{boquien2010a,boquien2010b,verley2010b}, which is of importance for the study of high redshift galaxies.
 \item The color trends of the warm and the cold dust show that they are heated by different sources. At higher SFR, the warm dust temperature seems to be driven by star formation. As star formation decreases, the temperature is increasingly driven by another component, most likely the evolved stellar populations. The cold dust temperature seems to be driven by the old stellar population, with a tight correlation with the local stellar mass.
\end{enumerate}

This paper lays the basis of additional studies in the context of the HERM33ES project which will go in much more detail on some of the points elaborated here. Upcoming papers will provide further insight on the PAHs emission (Rosolowsky et al., in preparation), the local dust temperature and mass (Xilouris et al., in preparation), the dust properties (Tabatabaei et al., in preparation), the HII regions (Rela\~no et al., in preparation), the local physical conditions using PACS spectroscopy (Mookerjea et al., 2011, submitted), etc.

\acknowledgments

This research has made use of the NASA/IPAC Extragalactic Database (NED) which is operated by the Jet Propulsion Laboratory, California Institute of Technology, under contract with the National Aeronautics and Space Administration.

\bibliographystyle{aa}
\bibliography{biblio}

\appendix
\section{Tables}

\begin{deluxetable}{cccccccccc}
\tablecolumns{9} \tablewidth{0pc}
\tablehead{\colhead{8.0}&\colhead{24}&\colhead{70}&\colhead{100}&\colhead{160}&\colhead{250}&\colhead{350}&\colhead{500}&b&\colhead{$\sigma$}}
\tabletypesize{\tiny}
\startdata
$ 0.842\pm 0.009$&--&--&--&--&--&--&--&$ 5.899\pm 0.285$&$ 0.087$\\
--&$ 0.816\pm 0.006$&--&--&--&--&--&--&$ 7.240\pm 0.200$&$ 0.065$\\
--&--&$ 0.915\pm 0.006$&--&--&--&--&--&$ 3.319\pm 0.206$&$ 0.059$\\
--&--&--&$ 0.962\pm 0.007$&--&--&--&--&$ 1.568\pm 0.251$&$ 0.067$\\
--&--&--&--&$ 1.027\pm 0.008$&--&--&--&$-0.574\pm 0.278$&$ 0.070$\\
--&--&--&--&--&$ 1.288\pm 0.013$&--&--&$-8.846\pm 0.419$&$ 0.084$\\
--&--&--&--&--&--&$ 1.391\pm 0.018$&--&$-11.673\pm 0.585$&$ 0.107$\\
--&--&--&--&--&--&--&$ 1.400\pm 0.023$&$-11.231\pm 0.757$&$ 0.134$\\
\enddata
\tablecomments{Coefficients of the fit of the combination of any number of Spitzer and Herschel bands from 8.0~$\mu$m to 500~$\mu$m to estimate the TIR luminosity. The last column give $\sigma$, the standard deviation of the TIR luminosity  around the best fit. The fitted relation is of the form: $\log L_{TIR}=\sum_{i=1}^n a_i \log S_i+b$, with $S_i$ in W~kpc$^{-2}$. The relation has been derived for brightnesses in the following range: $32.37<\log 8<34.45$, $31.98<\log 24<34.54$, $32.60<\log 70<34.95$, $33.29<\log 100<35.08$, $33.06<\log 160<34.79$, $32.65<\log 250<34.18$, $32.12<\log 350<33.65$,$31.60<\log 500<33.10$, where the brightness is in units of W~kpc$^{-2}$.\label{tab:fit-tir}}
\end{deluxetable}

\begin{deluxetable}{cccccccccc}
 \tablecolumns{9} \tablewidth{0pc}
\tablehead{\colhead{8.0}&\colhead{24}&\colhead{70}&\colhead{100}&\colhead{160}&\colhead{250}&\colhead{350}&\colhead{500}&b&\colhead{$\sigma$}}
\tabletypesize{\tiny}
\startdata
$ 0.835\pm 0.023$&--&--&--&--&--&--&--&$-30.340\pm 0.774$&$ 0.236$\\
--&$ 0.920\pm 0.015$&--&--&--&--&--&--&$-32.660\pm 0.500$&$ 0.163$\\
--&--&$ 1.038\pm 0.016$&--&--&--&--&--&$-37.309\pm 0.543$&$ 0.155$\\
--&--&--&$ 1.043\pm 0.021$&--&--&--&--&$-37.675\pm 0.700$&$ 0.188$\\
--&--&--&--&$ 1.065\pm 0.025$&--&--&--&$-38.364\pm 0.841$&$ 0.211$\\
--&--&--&--&--&$ 1.319\pm 0.033$&--&--&$-46.383\pm 1.109$&$ 0.222$\\
--&--&--&--&--&--&$ 1.430\pm 0.039$&--&$-49.439\pm 1.262$&$ 0.231$\\
--&--&--&--&--&--&--&$ 1.459\pm 0.042$&$-49.635\pm 1.359$&$ 0.241$\\
\enddata
\tablecomments{Coefficients of the fit of the combination of any number of Spitzer and Herschel bands from 8.0~$\mu$m to 500~$\mu$m to estimate the SFR in M$_\sun$~yr$^{-1}$~kpc$^{-2}$. The last column give $\sigma$, the standard deviation of the SFR around the best fit. The fitted relation is of the form: $\log SFR=\sum_{i=1}^n a_i \log S_i+b$, with $S_i$ in W~kpc$^{-2}$. The relation has been derived for brightnesses in the following range: $32.37<\log 8<34.45$, $31.98<\log 24<34.54$, $32.60<\log 70<34.95$, $33.29<\log 100<35.08$, $33.06<\log 160<34.79$, $32.65<\log 250<34.18$, $32.12<\log 350<33.65$,$31.60<\log 500<33.10$, where the brightness is in units of W~kpc$^{-2}$.\label{tab:fit-sfr}}
\end{deluxetable}

\end{document}